\documentclass[11pt]{article}
\usepackage{times}

\usepackage{amsmath,amsfonts,bm}

\def\eqref#1{equation~\ref{#1}}

\def\1{\bm{1}}

\DeclareMathAlphabet{\mathsfit}{\encodingdefault}{\sfdefault}{m}{sl}
\SetMathAlphabet{\mathsfit}{bold}{\encodingdefault}{\sfdefault}{bx}{n}

\usepackage{acl}
\usepackage{hyperref}
\usepackage{url}
\usepackage[utf8]{inputenc} 
\usepackage[T1]{fontenc}    
\usepackage{hyperref}       
\usepackage{url}            
\usepackage{booktabs}       
\usepackage{amsfonts}       
\usepackage{nicefrac}       
\usepackage{microtype}      
\usepackage{bm}
\usepackage{graphicx}
\usepackage{times}
\usepackage{latexsym}
\usepackage{mathtools}
\usepackage{amsmath}
\usepackage{pgfplots} 
\usepackage{booktabs}
\usepackage{xspace}
\usepackage{cleveref}
\usepackage{verbatim} 
\usepackage{amsmath,amsfonts,bm}
\usepackage{arydshln}
\usepackage{enumitem}
\usepackage{wrapfig}
\usepackage{selectp}
\usepackage{algorithm}
\usepackage{algorithmic}
\usepackage{times}
\usepackage{latexsym}
\usepackage[T1]{fontenc}
\usepackage[utf8]{inputenc}
\usepackage{microtype}
\usepackage{inconsolata}
\usepackage{url}
\usepackage{amsmath,amssymb}
\usepackage{amsfonts}
\usepackage{helvet}
\usepackage{courier}
\usepackage{url}
\usepackage{float,color}
\usepackage{times}
\usepackage{algorithm}
\usepackage{algorithmic}
\usepackage{float}
\usepackage{pdfpages}
\usepackage[commandnameprefix=always]{changes}
\usepackage{epsfig}
\usepackage{stfloats}
\usepackage{url}
\usepackage{diagbox}
\usepackage{subfigure}
\usepackage{epstopdf}
\usepackage{multicol}
\usepackage{makecell}
\usepackage{amsmath}
\usepackage{longtable}
\usepackage{dsfont,amsfonts,color}
\usepackage{multirow}
\usepackage{booktabs}
\usepackage{colortbl}
\PassOptionsToPackage{table}{xcolor}
\usepackage{xr}
\usepackage{tablefootnote}
\usepackage{tabularx}
\usepackage{listings}
\usepackage{wrapfig}
\usepackage{algorithm}
\usepackage{amsthm}
\usepackage{enumitem}
\usepackage[subfigure]{tocloft}

\usepackage{todonotes} 

\usepackage{float}

\newtheorem{lemma}{Lemma}[section]
\newtheorem{theorem}[lemma]{Theorem}

\newtheorem{definition}[lemma]{Definition}

\usepackage{titlesec}

\title{PrivacyRestore: Privacy-Preserving Inference in Large Language Models via Privacy Removal and Restoration}

\author{%
 Ziqian Zeng\thanks{Equal contribution.}\ \ \thanks{Corresponding author}\ \ \textsuperscript{\rm 1}, 
 Jianwei Wang\footnotemark[1]\ \ \textsuperscript{\rm 1}, 
 Junyao Yang\footnotemark[1]\ \ \textsuperscript{\rm 1}, \\
 \textbf{Zhengdong Lu}\textsuperscript{\rm 1},  \textbf{Haoran Li}\textsuperscript{\rm 3}, 
 \textbf{Huiping Zhuang}\textsuperscript{\rm 1},
 \textbf{Cen Chen}\textsuperscript{\rm 1,2}\\
\textsuperscript{\rm 1}South China University of Technology\\
\textsuperscript{\rm 2} Pazhou Laboratory, China\\
\textsuperscript{\rm 3} Hong Kong University of Science and Technology\\
{$\;\:$ \texttt{\href{mailto: zqzeng@scut.edu.cn}{zqzeng}@scut.edu.cn}},
{$\;\:$ \texttt{\href{mailto: wiwjwilliam@mail.scut.edu.cn}{wiwjwilliam}@mail.scut.edu.cn}}
}

\begin{document}

\maketitle
\begin{abstract}
The widespread usage of online Large Language Models (LLMs) inference services has raised significant privacy concerns about the potential exposure of private information in user inputs. 
Existing privacy protection methods for LLMs suffer from either insufficient privacy protection with performance degradation, or large inference time overhead. 
To address these limitations, we propose PrivacyRestore, a plug-and-play method to protect the privacy of user inputs during LLM inference for the client-server scenario. 
The server first trains restoration vectors for each privacy span type offline and then releases them to the clients.
During inference, the client aggregates restoration vectors of all privacy spans in the user query into a meta restoration vector which is later sent to the server to restore information.
Before transmission, the client removes all privacy spans in the user query and applies $d_\chi$-privacy mechanism to the meta vector for privacy protection.
We prove that our method can inherently prevent the linear growth of the privacy budget.
We conduct extensive experimental, covering the medical and legal domains, and demonstrate that PrivacyRestore effectively protects private information and maintains acceptable levels of performance and inference efficiency
\footnote{We release our code and datasets in \url{https://github.com/wjw136/PrivacyRestore}}
.
\end{abstract}

\section{Introduction}


Large language models (LLMs) have emerged as powerful tools across various domains \citep{chen2023meditron, wisdomInterrogatory, wu2023bloomberggpt}. 
However, the widespread use of online LLM inference services has raised significant privacy concerns.
When interacting with LLMs deployed on cloud platforms, users' inputs may contain sensitive data, such as medical records and legal case details.
Potential threats may arise when attackers intercept user queries during data transmission, and some advanced adversaries can even hack the cloud service provider. 
For example, in sensitive domains like medical diagnosis, if a user's input containing the user's personal protected health information (PHI), such as ``\textit{I was previously diagnosed with HIV, and lately I've been experiencing fever and diarrhea...}'' is disclosed to malicious attackers, it may cause privacy concerns. 
In this paper, we focus on protecting the private information contained in user inputs during LLM inference. 
In this setting, the client submits inputs to the server (also known as the service provider) and there is a risk that inputs might be disclosed by attackers.
Current methods for protecting user inputs can be mainly divided into two categories: Secure Multi-Party Computation (SMPC) and Differential Privacy (DP).
SMPC-based methods \citep{mao2022iron, li2023mpcformer, liang2024merge} utilize encryption protocols and algorithms to enable collaborative computation without revealing original data to others. 
However, SMPC methods have large inference time overheads, making them impractical for real-time applications \citep{hao2022iron}. 
DP based methods \citep{feyisetan2020privacy, feyisetan2019privacy, xu2020diff, bo2021er} apply $d_{\chi}$-privacy \citep{konstantinos2013broad, pazii2018local} to words and achieve word-level text-to-text privatization.
Nevertheless, DP-based methods inevitably degrade the performance of downstream tasks due to noise injection, which is known as the privacy-utility trade-off. 
Additionally, as the text length grows, word-level privatization will lead to significant performance degradation.
This phenomenon is known as the linear growth of the privacy budget in word-level privatization \cite{justus2022limit}.
Hence, there is a need to develop privacy-preserving methods which can effectively safeguard the privacy of user inputs while maintaining high-quality outputs, without incurring prohibitive computational costs.

We propose PrivacyRestore, which directly removes privacy spans in user inputs and restores private information via activation steering \citep{li2023iti,turner2023activation, hernandez2023inspecting} during model inference. 
Our method is based on two key phenomena:
\textbf{(a) Users' private information mostly consists of sensitive attributes and these attributes are commonly confined within specific contiguous token sequences, referred to as ``privacy spans''.} 
For eaxmple, in the context of healthcare domain, the private information may commonly refer to symptom descriptions. 
Consider a medical record which states ``\textit{I was previously diagnosed with HIV, and lately I've been experiencing fever and diarrhea...}'', ``\textit{HIV}" and "\textit{fever and diarrhea}'' should be protected as privacy spans.
Directly removing these privacy spans (symptom descriptions) can significantly hinder attackers from reconstructing or inferring private information and serves as an effective approach to preventing privacy leakage.
\textbf{(b) Most privacy spans are concentrated in a few majority categories, exhibiting a long-tailed distribution.}
For instance, in medical diagnosis applications, privacy spans typically relate to symptoms and disease descriptions. 
Most of the symptoms and disease descriptions appearing in user inputs are concentrated on high-frequency types, such as ``\textit{fever}'' and ``\textit{cold}''.
We have conducted experiments to demonstrate the long-tailed distribution of privacy spans, as detailed in Appendix \ref{app:long_dis}.

PrivacyRestore operates in two stages: the preparation stage and the inference stage. 
\textbf{In the preparation stage}, the server first identifies the attention heads where the activation steering occurs. 
Second, each privacy span type is encoded to a vector, known as the restoration vector. 
This stage is performed entirely offline on the server side.
Our method is plug-and-play, requiring only the restoration vectors to be trainable, while keeping the LLM frozen. 
Once training is complete, these restoration vectors will be released to the client side.
\textbf{In the inference stage}, according to the principle of ``Information Self-Determination Right'' \footnote{\url{https://en.wikipedia.org/wiki/Informational_self-determination}} \citep{ Mar2009Privacy, Van2009privacymind}
, the users are entitled to identify the privacy spans in their inputs by themselves.
After identification, a meta vector is constructed by estimating the importance of each privacy span and calculating a weighted sum of the corresponding restoration vectors.
Then, the user submits the incomplete input with the privacy spans removed, along with the meta vector, to the server. 
The server uses the meta vector to restore the removed privacy spans and generate high-quality outputs.



To prevent the leakage of privacy spans via the meta vector, $d_{\chi}$-privacy mechanisms are applied to the meta vector before transmission at the client side. 
By applying $d_{\chi}$-privacy to the meta vector instead of words, \textbf{our method inherently addresses the linear growth issue of privacy budget encountered in word-level privatization \citep{mattern2022differentially}}.
To further prevent privacy leakage through generated outputs, the server should employ sampling-based generation, enabling the output to be protected by the Exponential Mechanism \citep{Utpala2023LocallyDP, Mattern2022TheLO, Mc2007Mecha}.
Experimental results demonstrate that our method can effectively protect private information and maintain satisfactory performance and inference efficiency.
The contributions are summarized as follows:


\begin{itemize}[topsep=0.2em, partopsep=0em, itemsep=0.5em, parsep=0em, leftmargin=2em]
    \item We propose a plug-and-play privacy protection method that removes privacy spans in the input and restores private information via activation steering during inference. 
    \item We propose Attention-aware Weighted Aggregation to construct the meta vector and apply the $d_{\chi}$-privacy mechanism to the meta vector, inherently addressing the problem of the linear growth of privacy budget.
    \item We construct three datasets, covering the medical and legal fields, to evaluate our method. Experimental results demonstrate its capabilities of privacy protection. It also maintains acceptable performance and inference efficiency.
\end{itemize}

\section{Related Works}


In this section, we introduce the related works on user input protection methods, which are currently divided into two categories: SMPC-based methods and DP-based methods. 


\paragraph{SMPC-based methods.}

Secure Multi-Party Computation (SMPC) uses encryption algorithms to enable secure collaborative computations between the client and server, without revealing the original user inputs to the server.

However, SMPC incurs significant inference time overhead, rendering it impractical for real-time LLM applications \citep{li2023mpcformer, liang2024merge, mao2022iron, liu2023encrypt, zheng2023primer, kanav2023sigma, lu2024zero}.

\paragraph{DP-based methods.}

$d_\chi$-privacy mechanisms, a variant of DP, protect user inputs by injecting noise. However, this approach can lead to performance degradation. 

Additionally, $d_\chi$-privacy becomes less effective as input length increases, due to the linear growth of the privacy budget \citep{feyisetan2019privacy, justus2022limit, sai2023ldp, cyn2016cali, john2013local}.

Due to the limited space, a detailed introduction of the above works can be found in Appendix \ref{app:appendix_related_work}.

\section{Threat Model}
\label{sec:threat_model}
We consider a threat model involving two parties: a server that holds the LLM weights and a client holds user inputs containing privacy spans. 
Privacy span is defined as \textbf{a contiguous token sequence that contains private information} in user inputs, and should be identified by the user itself according to the principle of ``Information Self-Determination Right''.
The server provides services through an API while maintaining the confidentiality of the LLM weights.
Adversaries may intercept privacy spans when users submit their inputs via the API.
Relying solely on encryption algorithms is insufficient to prevent privacy leakage, as encrypted inputs will be decrypted on the server and the server itself may be vulnerable.
Some advanced adversaries can even hack the server to steal those decrypted user inputs easily \citep{law2024unit, Bill2024Ann}.
Therefore, our goal is to protect the privacy spans that exist in user inputs from attackers who are capable of stealing user privacy during transmission or even hacking the server to steal user privacy.


\section{Methodology}
\label{sec:method}
\begin{figure*}[!ht]
\centering
\includegraphics[width=1\textwidth]{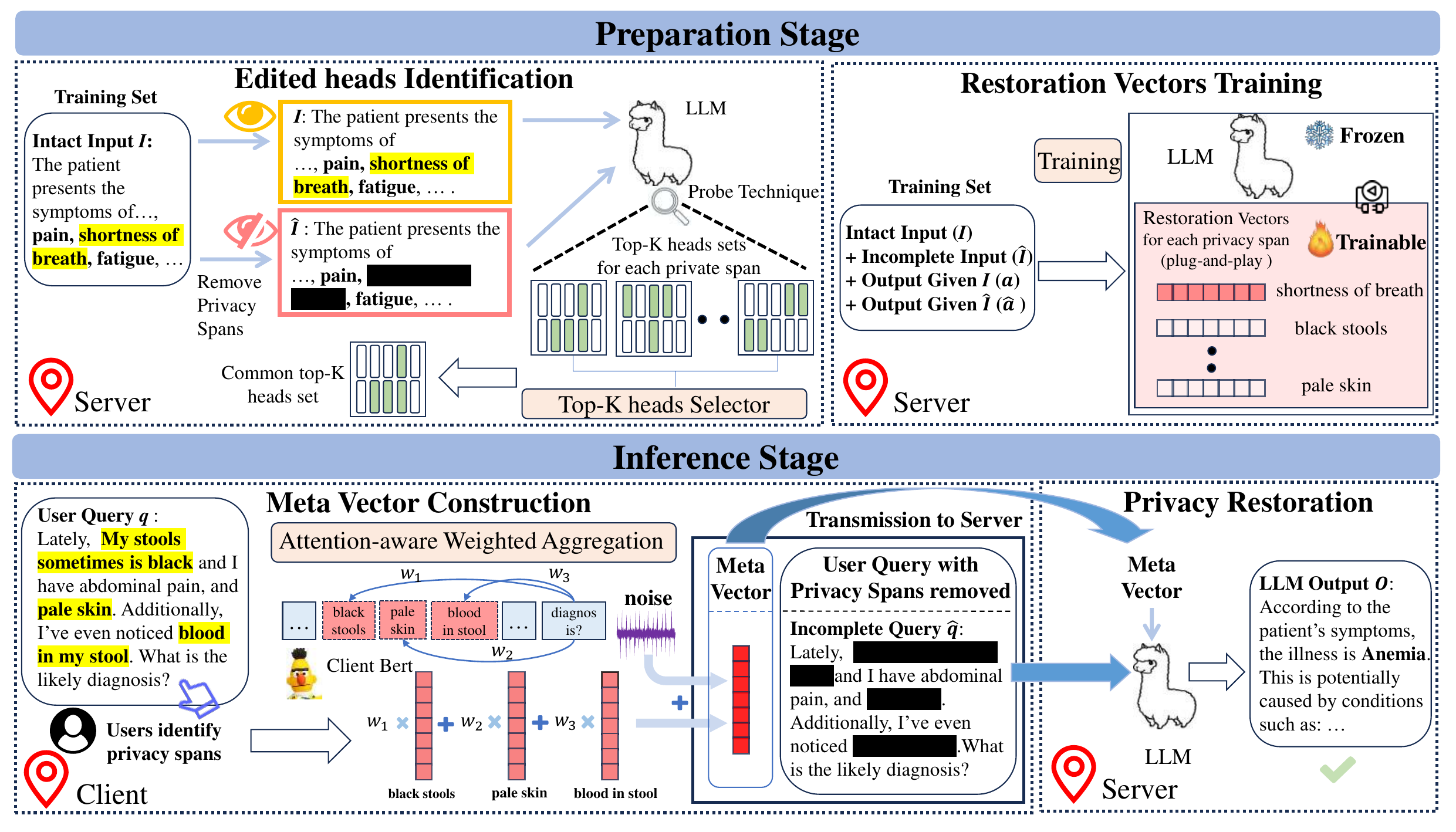}
\vspace{-15pt}
\caption{
The PrivacyRestore consists of two stages.
(1) \textbf{Preparation Stage.} 
This stage aims to identify the edited heads and train the restoration vectors.
We provide a more detailed training set example in Figure \ref{fig:training_samples}.
(2) \textbf{Inference Stage.} 
In this stage, the client constructs a meta vector.
The server uses the meta vector to restore information during inference on the incomplete query.
}
\vspace{-1.5em}
\label{fig:algorithm}
\end{figure*}

PrivacyRestore operates in two stages, i.e., the preparation stage and the inference stage, as shown in Figure \ref{fig:algorithm}:

(1) \textbf{Preparation stage}: This stage takes place only on the server. 
Considering the long-tailed distribution of privacy spans, we predefine a core set of privacy span types that covers the majority of them.
Next, we \textbf{identify the edited attention heads} required for activation steering during the inference stage.
Finally, we \textbf{train the restoration vector} for each privacy span type in the predefined core set.
After training, all these vectors are released to the clients. 
The preparation stage is conducted offline, prior to the server beginning to offer its services.

(2) \textbf{Inference stage}: This stage involves collaboration between the client and server.
According to the principle of ``Information Self-Determination Right'', the users should identify all privacy spans in their queries by themselves.
Then, the client removes all these privacy spans from the queries for privacy protection. For restoration, the client \textbf{constructs a meta vector} according to the removed privacy spans and applies $d_\chi$-privacy to the meta vector to prevent privacy leakage.
The meta vector, along with the incomplete queries with privacy spans removed, are sent to the server.
The server performs inference on the incomplete input and \textbf{restores information} using the meta vector through activation steering.


The preparation and inference stages descriptions are provided in \S \ref{sec:prepare} and \S \ref{sec:inference}, respectively.
All notation definitions are shown in Appendix \ref{sec:notations}. 
Backgrounds about the $d_\chi$-privacy mechanism and activation steering are shown in Appendix \ref{app:method_preliminary}.

\setlength{\textfloatsep}{10pt}

\subsection{Preparation Stage}
\label{sec:prepare}
\paragraph*{Edited Heads Identification.} 
As indicated by activation steering methods \citep{li2023iti, chen2024truth}, modifying all attention heads in LLMs will degrade overall performance. 
Inspired by this, we aim to identify the attention heads most relevant to privacy spans. 
 

As shown in the upper part of Figure \ref{fig:algorithm}, we firstly utilize the probe technique \citep{alain2016understanding, tenney2019bert, belinkov2022probing} to identify the most relevant attention heads for each privacy span type. 
We train a binary classifier for each head, tailored to the privacy span type $c$, as the probe. 
A probe with higher accuracy indicates a stronger correlation between the head $h$ and the privacy span type $c$.
Therefore, we select the top $K$ attention heads with the highest accuracies for each privacy span type $c$ in the predefined core set $\mathcal{C}$.
Using different top-K head sets for different privacy span types may suffer the risk of privacy leakage, as an attacker could infer the presence of a specific privacy span type based on the characteristics of top-K head set. 
Hence, we propose a \textbf{Common Top-K Selector} to combine all different top-K head sets to construct a common top-K head set $\mathcal{H}_k$ as the edited head set.
To achieve this, we calculate the average score of each head across all privacy span types in $\mathcal{C}$, selecting the highest $K$ heads to construct the common set.
A head receives a positive score if it appears in the top-K head set of a privacy span type $c$. 
The score is related to the accuracy of the probe associated with the head. 
\vspace{-0.1em}
\paragraph{Restoration Vectors Training.}
\label{sec:res_corr}
After identifying the edited heads set $\mathcal{H}_k$, the next step is to train the restoration vectors for each privacy span type in the predefined core set $\mathcal{C}$. 

For each privacy span type $c \in \mathcal{C}$, there is a trainable restoration vector $r_{h}^{c}$  for each head $h$ in the common top-K heads set $\mathcal{H}_k$. 
The restoration vectors constitute the only trainable parameters $\Theta$ in our method, while the LLM weights held by the server remain fixed.
Therefore, our method is plug-and-play and parameter-efficient for training.

We fine-tune the restoration vectors using the ORPO loss proposed by \citet{orpo}, which integrates the supervised fine-tuning process and the preference alignment process.
This loss function can guide the model in generating better answers.
In our method, we use ORPO loss to train the restoration vectors $\Theta$, ensuring that the outputs generated from inputs without privacy spans and restored using the corresponding restoration vectors, can closely resemble those generated from the intact inputs.
More details of the probe technique, the common top-K head set construction and restoration vectors training process are provided in Appendix \ref{app:probe_techniques}, Appendix \ref{sec:selector} and Appendix \ref{sec:details_train}, respectively.
After restoration vectors training, the server will release all restoration vectors to clients.

\subsection{Inference Stage}
\label{sec:inference}
\paragraph{Meta Vector Construction.} 
According to the principle of ``Information Self-Determination Right'', users should identify the privacy spans in their input by themselves, because the definition of privacy varies from person to person.
For each privacy span, the client employs a lightweight model (e.g. BERT \cite{devlin2019bert}) to classify it into a specific type within the predefined privacy span type set $\mathcal{C}$.
For example, the privacy span ``\textit{My stools sometimes is black}'' will be classified into the predefined privacy span type \textit{black stools}. 
Due to the long-tailed distribution of privacy spans, our predefined type set can cover the majority of privacy spans. 
Even when encountering privacy spans of out-of-set types, classifying these rare spans into the types of the predefined set can still be effective, as shown in \S \ref{sec:out_of_set}.

Then the client should aggregate those restoration vectors corresponding to the privacy spans into a single meta vector.
Transmitting a singular meta vector enhances privacy protection compared to multiple vectors, as it prevents potential information leakage regarding the quantity of privacy-sensitive segments.

However, Equal Weighted Aggregation (EWA) may weaken the influence of critical privacy spans and amplify the effect of irrelevant ones. 
Therefore, we propose a novel method called \textbf{Attention-aware Weighted Aggregation (AWA)} which estimates a weight for each privacy span and then takes the weighted sum of corresponding restoration vectors as the meta vector. 
Given computational constraints on client devices, we employ a lightweight BERT model to evaluate privacy span significance by calculating the mean attention score $w_s$ across all attention heads and tokens within the input query. This metric quantifies the relative importance of each privacy-related span.

Considering that each privacy span type $c$ will have multiple restoration vectors $r_h^c$ across all edited heads in $\mathcal{H}_k$, we first concatenate these restoration vectors from multiple heads to form $r^c$ for privacy span type $c$. 
Then, we compute the meta vector $\mathcal{R}$ by calculating the weighted sum of the restoration vector $r^c$, normalizing the summary, and adding noise $\mathcal{N}$ for privacy protection.
The process is formulated as follows:
\begin{eqnarray}
     r^c &=& \text{Concat}(r_1^c, r_2^c, ..., r_h^c), \\
     \label{eq:orgianl_meta_vec}
     Z &=& \frac{\sum_{s\in\mathcal{S}_q} w_s \cdot r^{c}}{||\sum_{s\in\mathcal{S}_q} w_s \cdot r^{c}||_2}, \\
     \label{eq:add_noise}
     \mathcal{R} &=&  Z + \mathcal{N},
\end{eqnarray}
where $s$ represents the privacy span of type $c$, $\mathcal{S}_q$ denotes all privacy spans in the user query $q$ and $Z$ represents the normalization of the weighted sum, which can also be viewed as the meta vector without protection. 
The injected noise $\mathcal{N}$ is sampling from the distribution $p(\mathcal{N}) \propto \exp(-\epsilon\|\mathcal{N}\|)$, to achieve the $d_\chi$-privacy mechanism, where $\epsilon$ is the privacy hyperparameter \citep{feyisetan2020privacy}.

After construction, the meta vector $\mathcal{R}$ and the incomplete query $\hat{q}$ (with privacy spans removed) are transmitted to the server for inference.

\paragraph{Privacy Restoration.}
We utilize the meta vector $\mathcal{R}$ to restore the information in the removed privacy spans during inference, as illustrated in the lower right part of Figure \ref{fig:algorithm}. This operation is conducted on the server side. 

Following activation steering methods \citep{li2023iti, chen2024truth}, we apply the meta vector to the outputs of the edited attention heads to achieve restoration.
Let ${\textbf{u}}_{h}$ represent the hidden state of the last token on head $h$ given the incomplete user query $\hat{q}$ and $\mathcal{R}_{h}$ denotes a part of the meta vector 
$\mathcal{R}$ for head $h$.
Then the hidden state of the last token on head $h$ after restoration, denoted as $\bar{\textbf{u}}_{h}$, can be computed by:
\begin{eqnarray}
\label{eq:restore}
     \bar{\textbf{u}}_{h} = {\textbf{u}}_{h} + ||\textbf{u}_h||_2\cdot\mathcal{R}_{h},\  \forall h \in \mathcal{H}_k. 
\end{eqnarray}
During inference, if a head belongs to the common top-K heads set $\mathcal{H}_k$, its hidden state should be modified using Eq \ref{eq:restore}. 
To prevent privacy leakage from the generated output, we employ sampling-based generation, which is protected by the Exponential Mechanism \citep{Utpala2023LocallyDP}.

\section{Privacy Guarantee Analysis}
\label{sec:priana}
In this section, we analyze the privacy guarantees and privacy budget of PrivacyRestore.

Our approach transmits only a privacy-free incomplete query and a meta vector secured by $d_\chi$-privacy mechanism.
Therefore, even if attackers steal both the incomplete user query and the meta vector during transmission or even hack the server, they still cannot infer any user privacy.
Furthermore, the confidentiality of the server’s LLM parameters and edited head set $\mathcal{H}_k$ prevents attackers from reconstructing the generation process using intercepted elements, ensuring robust security against privacy breaches.
Then we analyze the privacy budget of our method, as follows:
\begingroup
\setlength{\topsep}{0.5em} 
\begin{theorem}
\label{theo_2}
\textit{PrivacyRestore fulfills $d_{\chi}$-privacy and provides a privacy budget of $2\epsilon$, where $\epsilon$ denotes privacy hyperparameter.
The privacy budget of PrivacyRestore is independent of the length of the protected text. }
\end{theorem}
 
Pointed by \citet{justus2022limit}, directly applying $d_\chi$-privacy mechanism to all tokens in the user query, for privacy protection, suffers from the linear growth problem of privacy budget.
In contrast, our method ensures that the privacy budget remains constant at $2\epsilon$, independent of the length of protected text.
We also provide empirical evidence to demonstrate that our approach effectively addresses the linear growth problem of the privacy budget encountered in $d_\chi$-privacy in Section \ref{sec:longer_q}.
Detailed proof of Theorem \ref{theo_2} is provided in Appendix \ref{sec:fur_theo_2}.

\section{Experiments}
\subsection{Experiments Setup}
\label{sec:setup}
\paragraph{Datasets.} 
To evaluate our method, we construct three privacy-preserving datasets covering the medical and legal domains: \textbf{Pri-DDXPlus}, \textbf{Pri-NLICE}, and \textbf{Pri-SLJA}.
The detailed process of dataset construction and statistical information can be found in Appendix \ref{app:dataset}.

\vspace{-4px}
\paragraph{Metrics.}
The evaluation assesses both performance and inference efficiency. 
For performance evaluation, we use \textbf{MC1/MC2} \citep{zhang2024truthx}, ROUGE-L \citep{lin-2004-rouge}, and \textbf{LLM-Judge (LLM-J)} \citep{Lia2023judging} metrics. 
For inference efficiency, we use the \textbf{Throughput (TP)} metric.
The details of these metrics and their corresponding calculation processes are provided in Appendix \ref{app:eva_metrics}.

\vspace{-4px}
\paragraph{Compared Methods.}
To demonstrate the effectiveness of our method, we compare our model with the following baselines:
\textbf{No Protection},
\textbf{No Restoration},
\textbf{${d_\chi}$-privacy} \citep{feyisetan2020privacy}, \textbf{${d_\chi}$-privacy on privacy spans} and \textbf{Paraphrase} \cite{justus2022limit, sai2023ldp}.
A detailed introduction to these baseline methods is provided in Appendix \ref{app:com_method}.

\vspace{-4px}
\paragraph{Settings of Privacy Hyperparameters.}
The hyperparameters related to privacy protection strength are $\epsilon$ for $d_{\chi}$-privacy (on privacy spans) and PrivacyRestore, and $\tau$ for paraphrase. 
For a fair comparison, we ensure all methods are under the same privacy budget.
We show the calculation process of determining values of $\epsilon$ and $\tau$ for different methods on different datasets in Appendix \ref{app:para}. 

\subsection{Main Results}
\begin{table*}
    \centering
    \def\arraystretch{1}
    \resizebox{0.92\textwidth}{!}{
    \begin{tabular}{ l l |  ccccc  }
    \toprule
        \textbf{Datasets} & \textbf{Methods}  & \textbf{MC1} $\uparrow$ & \textbf{MC2} $\uparrow$ & \textbf{ROUGE-L} $\uparrow$ & \textbf{LLM-J} $\uparrow$ & \textbf{TP} $\uparrow$ \\
        \cmidrule(lr){1-7} 
        \multirow{4}{*}{Pri-DDXPlus}
        & No Restoration (lower bound) &
        33.57$_{\pm0.00}$ & 32.49$_{\pm0.01}$ & 25.19$_{\pm0.43}$ & 3.21$_{\pm0.01}$ & 40.86$_{\pm0.01}$ \\
        & No Protection (upper bound) &
        64.88$_{\pm0.01}$ & 61.48$_{\pm0.03}$ & 100.00$_{\pm0.00}$ & 5.58$_{\pm0.03}$  & \textbf{41.08}$_{\pm0.09}$\\
        & $d_\chi$-privacy &
        28.79$_{\pm0.02}$ & 30.26$_{\pm0.01}$ & 17.97$_{\pm0.00}$ & 1.17$_{\pm0.00}$ & 37.45$_{\pm0.01}$ \\
        & $d_\chi$-privacy on privacy spans &
        44.71$_{\pm0.29}$ & 42.36$_{\pm0.00}$ & \textbf{29.17$_{\pm0.04}$} & 3.31$_{\pm0.00}$ & 33.21$_{\pm0.00}$ \\
        & Paraphrase &
        27.92$_{\pm0.56}$ & 28.56$_{\pm0.07}$ & 18.04$_{\pm0.01}$ & 1.23$_{\pm0.00}$ & 35.42$_{\pm0.67}$ \\
        \rowcolor{lightgray!45}
        \cellcolor{white} & \textbf{PrivacyRestore} & \textbf{62.97$_{\pm0.00}$} & \textbf{60.19$_{\pm0.00}$} &
        27.24$_{\pm0.26}$ & \textbf{4.47$_{\pm0.00}$} & 26.09$_{\pm0.08}$ \\

        \cmidrule(lr){2-7} 
        \multirow{4}{*}{Pri-NLICE}
        & No Restoration (lower bound) & 27.07$_{\pm1.98}$ & 28.63$_{\pm2.23}$ & 16.90$_{\pm0.51}$ & 1.61$_{\pm0.03}$& 41.08$_{\pm0.01}$ \\
        & No Protection (upper bound) & 80.30$_{\pm0.38}$ & 77.60$_{\pm1.23}$ & 100.00$_{\pm0.00}$ & 5.90$_{\pm0.04}$ & \textbf{41.44}$_{\pm0.04}$ \\
        
        & $d_\chi$-privacy & 29.08$_{\pm0.00}$ & 29.72$_{\pm0.00}$ & 15.68$_{\pm0.02}$ & 1.41$_{\pm0.00}$ & 38.30$_{\pm0.00}$ \\
        & $d_\chi$-privacy on privacy spans & 30.00$_{\pm0.09}$ & 31.46$_{\pm0.00}$ & 22.97$_{\pm0.00}$ & 3.01$_{\pm0.00}$ & 35.73$_{\pm0.57}$  \\
        & Paraphrase & 28.46$_{\pm0.02}$ & 29.15$_{\pm0.03}$ & 16.15$_{\pm0.01}$ & 1.62$_{\pm0.00}$ & 37.22$_{\pm0.07}$ \\
        \rowcolor{lightgray!45}
        \cellcolor{white} & \textbf{PrivacyRestore} & \textbf{62.23$_{\pm1.70}$} & \textbf{57.94$_{\pm0.09}$} &
        \textbf{24.42$_{\pm0.81}$} & \textbf{3.67$_{\pm0.01}$} & 32.33$_{\pm0.01}$ \\

        \cmidrule(lr){2-7} 
        \multirow{4}{*}{Pri-SLJA}
        & No Restoration (lower bound) & 24.92$_{\pm0.98}$ & 25.97$_{\pm1.12}$ & 31.02$_{\pm0.16}$ & 4.43$_{\pm0.01}$ & 39.14$_{\pm0.09}$ \\
        & No Protection (upper bound) &
        69.57$_{\pm0.61}$ & 67.58$_{\pm0.43}$ & 100.00$_{\pm0.00}$ & 5.44$_{\pm0.03}$ & \textbf{39.49}$_{\pm0.13}$ \\

        & $d_\chi$-privacy &
        16.66$_{\pm0.37}$ & 17.57$_{\pm0.04}$ & 23.35$_{\pm0.00}$ & 2.08$_{\pm0.00}$ &
        36.83$_{\pm0.03}$ \\
        & $d_\chi$-privacy on privacy spans & 24.23$_{\pm1.69}$ & 26.63$_{\pm0.67}$ & \textbf{40.10$_{\pm0.00}$} & 4.54$_{\pm0.00}$ & 36.16$_{\pm0.00}$ \\
        & Paraphrase & 16.21$_{\pm0.02}$ & 17.52$_{\pm0.02}$ & 24.90$_{\pm0.01}$ & 2.07$_{\pm0.01}$ & 31.31$_{\pm0.05}$ \\
        \rowcolor{lightgray!45}
        \cellcolor{white} & \textbf{PrivacyRestore} & \textbf{35.47$_{\pm1.48}$} & \textbf{35.41$_{\pm0.64}$} &
        37.56$_{\pm0.06}$ & \textbf{5.25$_{\pm0.00}$} & 30.73$_{\pm0.04}$ \\
        \midrule
    \end{tabular}
    }
    \caption{
    Comparison of the performance and the inference efficiency between PrivacyRestore and other baselines across three privacy-preserving datasets.
    All experiments are conducted over 3 runs, with the average results and variances reported. 
    The best results are highlighted in \textbf{bold}.
    \vspace{-1em}
    }
    \label{tbl:main}
\end{table*}

As shown in Table \ref{tbl:main}, we evaluate the performance and inference efficiency of PrivacyRestore and other compared methods across three privacy-preserving datasets.
Compared to $d_\chi$-privacy and paraphrase, $d_\chi$-privacy on privacy spans solely apply $d_\chi$-privacy mechanism to those privacy spans and achieves higher scores in MC1/2, ROUGE-L and LLM-J.
The possible reason for this is that both $d_\chi$-privacy and paraphrase operate on the entire user input, instead of specific privacy spans. 
Injecting noise into the entire input creates larger disturbances during inference compared to only corrupting a limited number of privacy spans.


PrivacyRestore achieves best scores in MC1/2 and LLM-J compared to other privacy-preserving methods. 
In terms of the ROUGE-L evaluation metric, PrivacyRestore achieve the best result in Pri-NLICE while ranking second in the other two datasets. 
This discrepancy likely stems from ROUGE-L’s dependence on $n$-gram overlap between the reference text and the generated output, which does not fully reflect the quality of generated outputs. 
As demonstrated by the examples in Figure \ref{fig:example} and Appendix \ref{app:example}, PrivacyRestore often generates outputs with different sentence structures while still providing accurate answers. 
Consequently, our method achieves slightly lower ROUGE-L scores but significantly higher LLM-J scores compared to $d_\chi$-privacy on privacy spans. Furthermore, the ROUGE-L metric displays larger variance than the LLM-J metric, potentially due to its sensitivity to expression rather than the underlying meaning of the generated output.
Shown in Table \ref{tbl:main}, No Protection servers as the performance \textbf{upper bound} for all privacy-preserving methods while No Restoration servers as the performance \textbf{lower bound}.
Our method significantly outperforms No Restoration and is even comparable to No Protection, strongly validating the effectiveness of our approach.

Although PrivacyRestore incurs slight latency from client-side privacy span identification and meta-vector construction, its throughtput attain nearly 70\% of the best results, which is acceptable.
We provide further analysis and additional experimental results in Appendix~\ref{app:inference_efficiency}.


\subsection{Empirical Privacy Protection Results}
\label{sec:attack}
In this section, we implement attack methods to empirically show that our approach offers superior privacy protection compared to baselines, both for user inputs and model outputs.
\paragraph{Privacy Protection Evaluation on Inputs.}
\label{sec:hyper_epi} 
\label{sec:pri_protect_eval}
In this section, we implement the embedding inverse attack \citep{hao2023sentence, john2023text} and attribute inference attack \citep{hao2022you} to attack the inputs of our method and other baselines, including the meta vector and the privacy-free incomplete user query.
As shown in Figure \ref{fig:different_epislon}, as the privacy budget increases, the privacy protection capability of all privacy-preserving methods decreases. 
However, PrivacyRestore consistently outperforms others across all privacy budgets, as indicated by its lower ROUGE-L and F1 scores.


\vspace{-0.4em}
\begin{figure}[!htbp]
  \centering
\raggedleft
    \begin{minipage}{0.90\columnwidth}
    
    
    \includegraphics[width=1\textwidth]{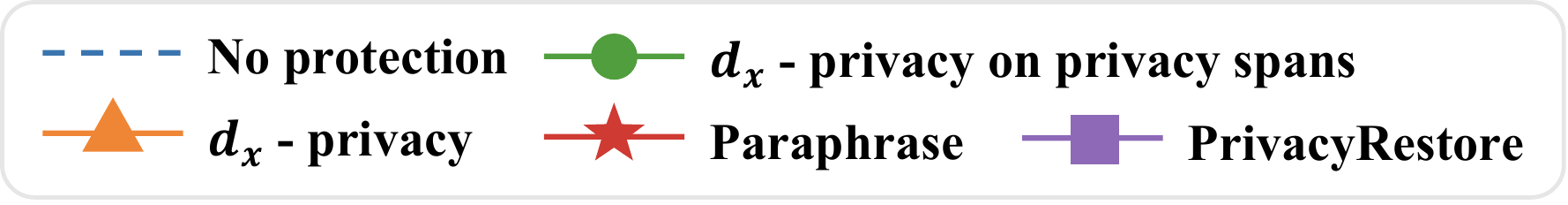}
    \label{fig:grow_legend}
  \end{minipage}
  \vspace{-2.2em}
  \vskip\baselineskip 
  \begin{minipage}{0.22\textwidth}
    \centering
    \subfigure[\centering Embedding Inverse Attack] {
     \label{fig:grow_pi}     
    \includegraphics[width=1\columnwidth]{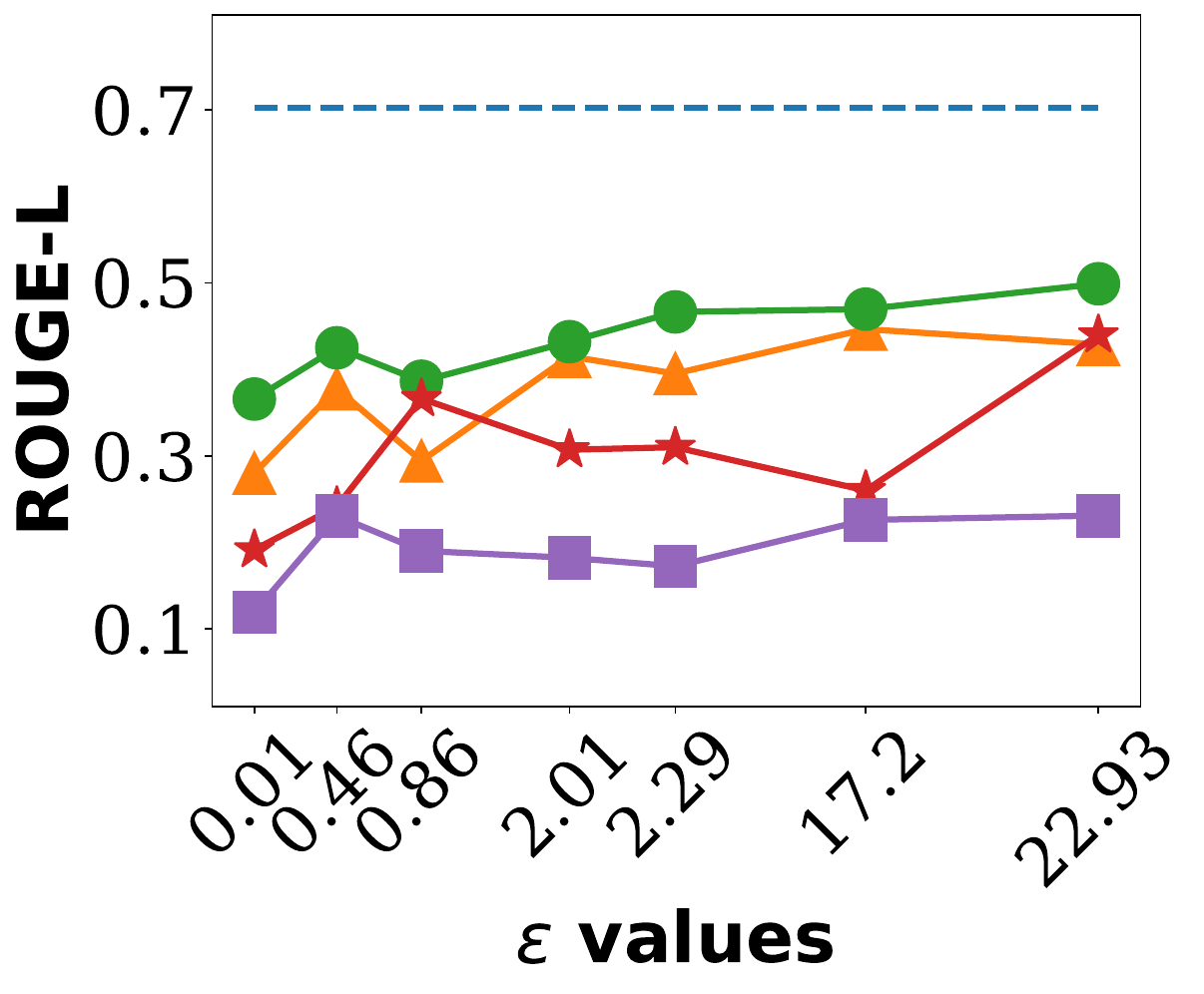}  
    }    
  \end{minipage}%
  \hfill
  \begin{minipage}{0.22\textwidth}
    \centering
    \subfigure[Attribute Inference Attack] {
     \label{fig:grow_ai}     
    \includegraphics[width=1\columnwidth]{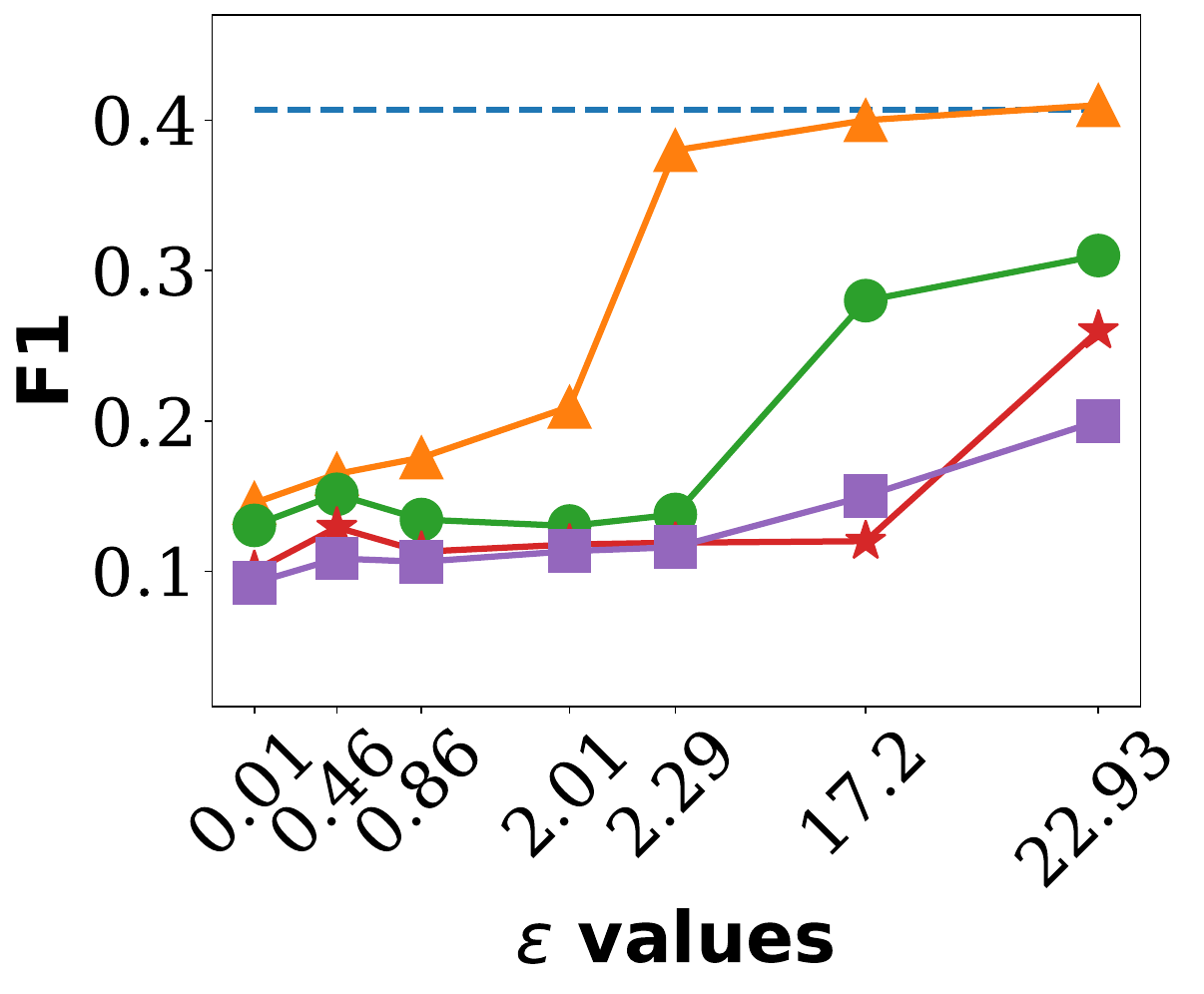}  
    }  
  \end{minipage}
  \vspace{-0.75em}
  \caption{Results of embedding inverse attack and attribute inference attack for all baselines under different privacy hyperparameters $\epsilon$ on Pri-DDXPlus.}
  \label{fig:different_epislon}
\end{figure}




\paragraph{Privacy Protection Evaluation on Outputs.}
We use the sampling-based method to generate the outputs on the server. 
As demonstrated by \citet{Utpala2023LocallyDP, justus2022limit}, sampling-based generation satisfies the Exponential Mechanism \citep{Mc2007Mecha}, which can effectively prevent privacy leakage from the generated outputs.
We implement the embedding inversion and attribute inference attacks for the generated outputs under various generation temperatures and also count the occurrence of privacy spans in the outputs.
As shown in Table \ref{table:output_privacy}, the attack performance remains consistently low, demonstrating that sampling-based generation effectively prevents privacy leakage from the generated outputs. 
Implementation details of these attack methods can be found in Appendix \ref{app:detail_PP_eval} and Appendix \ref{sec:output_privacy_spans}.

We also provide a biref theoretical proof of our method's output protection in Appendix~\ref{app:proof_output_pro}.

\begin{table}[!htbp]
\centering
\def\arraystretch{1.1} 
\resizebox{0.48\textwidth}{!}{
\begin{tabular}{l ccccc}
\toprule
 \textbf{Temperature} & \textbf{0.75} & \textbf{1.0} & \textbf{1.25} & \textbf{1.5}& \textbf{1.75} \\
\midrule
EIA(ROUGE-L) & 0.037 &  0.038& 0.035& 0.035 &0.037\\ 
AIA(F1) & 0.096 &  0.097& 0.092& 0.092&0.097 \\ 
Occurrence& 0.031 &  0.030& 0.030& 0.029&0.031 \\ 

\bottomrule
\end{tabular}
}   
\caption{Analysis of output privacy leakage from outputs on Pri-DDXPlus dataset. \textbf{EIA} denotes embedding inverse attack. \textbf{AIA} indicates attribute inference attack. \textbf{Occurrence} metric directly counts the frequency of privacy spans in the generated output. 
We primarily use a temperature of 1.0 during generation in the other experiments.}
\label{table:output_privacy}
\end{table}


\subsection{Privacy Protection for Long Text}
\label{sec:longer_q}
In this section, we implement attack methods for both the $d_\chi$-privacy baseline and our PrivacyRestore approach under varying protected text lengths, illustrating robust privacy and addressing the linear growth of the privacy budget in $d_\chi$-privacy.

For \textbf{$d_\chi$-privacy}, we randomly select a proportion of tokens as $d_\chi$-privacy Percentage to protect—higher percentages yield longer protected text. 
Shown in Figures \ref{fig:pi_dx} and \ref{fig:at_dx}, both prompt injection and attribute inference attacks are implemented with attack performance increases with the percentage.
It is caused by the linear growth problem of the privacy budget encountered in $d_\chi$-privacy, as raised by \citet{justus2022limit}.

For \textbf{PrivacyRestore}, a proportion of privacy spans is selected for protection, defined by the Privacy Span Ratio $\alpha$, with larger $\alpha$ indicating more spans. 
Shown in Figures \ref{fig:ea_pr} and \ref{fig:ai_pr}, aside from the embedding inverse attack on the Pri-NLICE dataset, attack performance remains stable across different $\alpha$ values. 
These results confirm that our method provides strong privacy protection, even as text length increases.

\begin{figure*}[!htbp]
    \centering
    \begin{minipage}{0.60\textwidth}
    
    
    \includegraphics[width=1\textwidth]{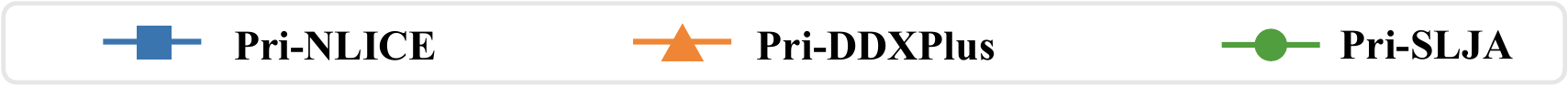}
    \label{fig:ratio_legend}
  \end{minipage}
  \vspace{-3em}
  \vskip\baselineskip 
    \subfigure[Prompt Injection Attack]{
        \includegraphics[width=0.22\textwidth]{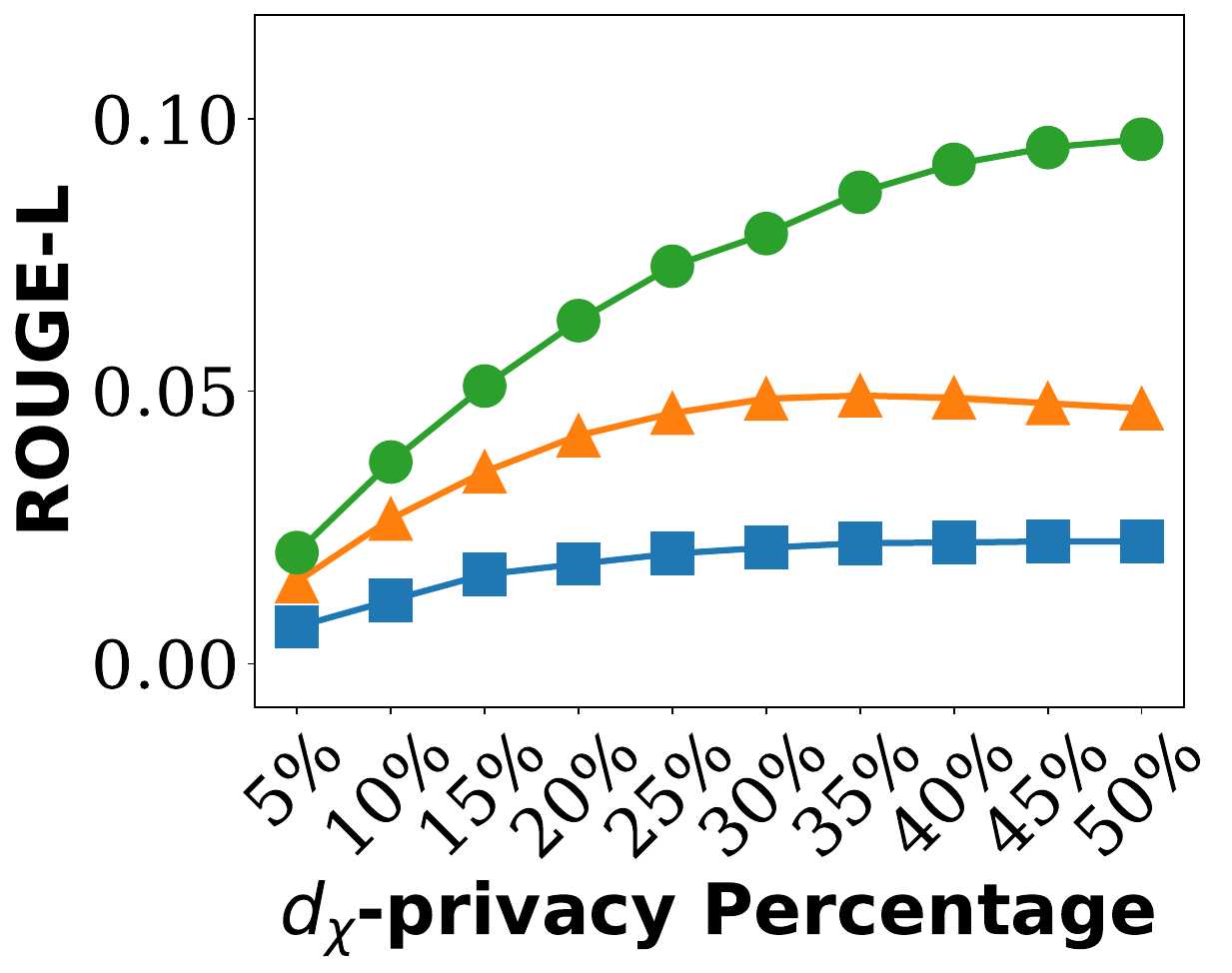}
        \label{fig:pi_dx}
    }
    \hfill
    \subfigure[Attribute Inference Attack]{
        \includegraphics[width=0.22\textwidth]{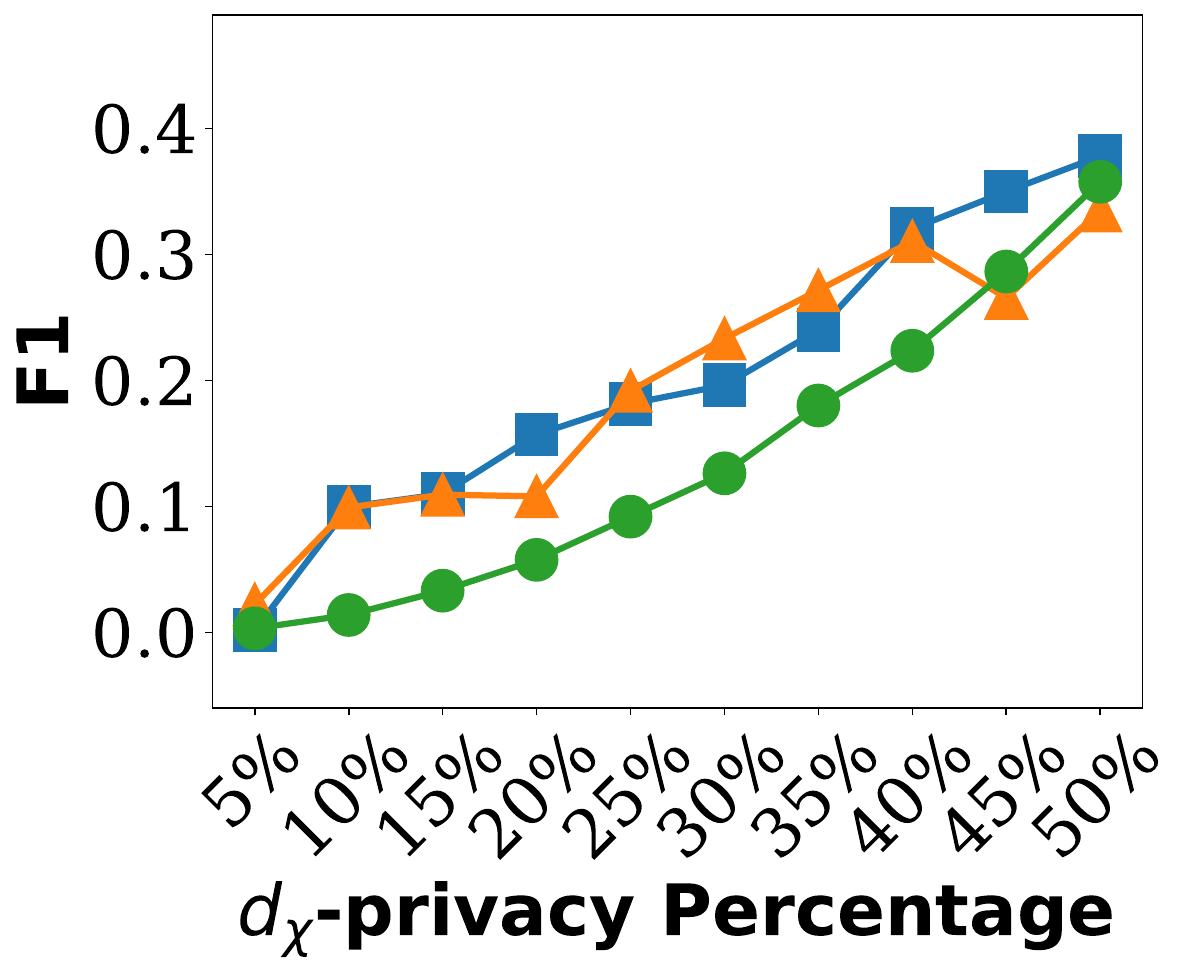}
        \label{fig:at_dx}
    }
    \hfill
    \subfigure[Embedding Inverse Attack]{
        \includegraphics[width=0.22\textwidth]{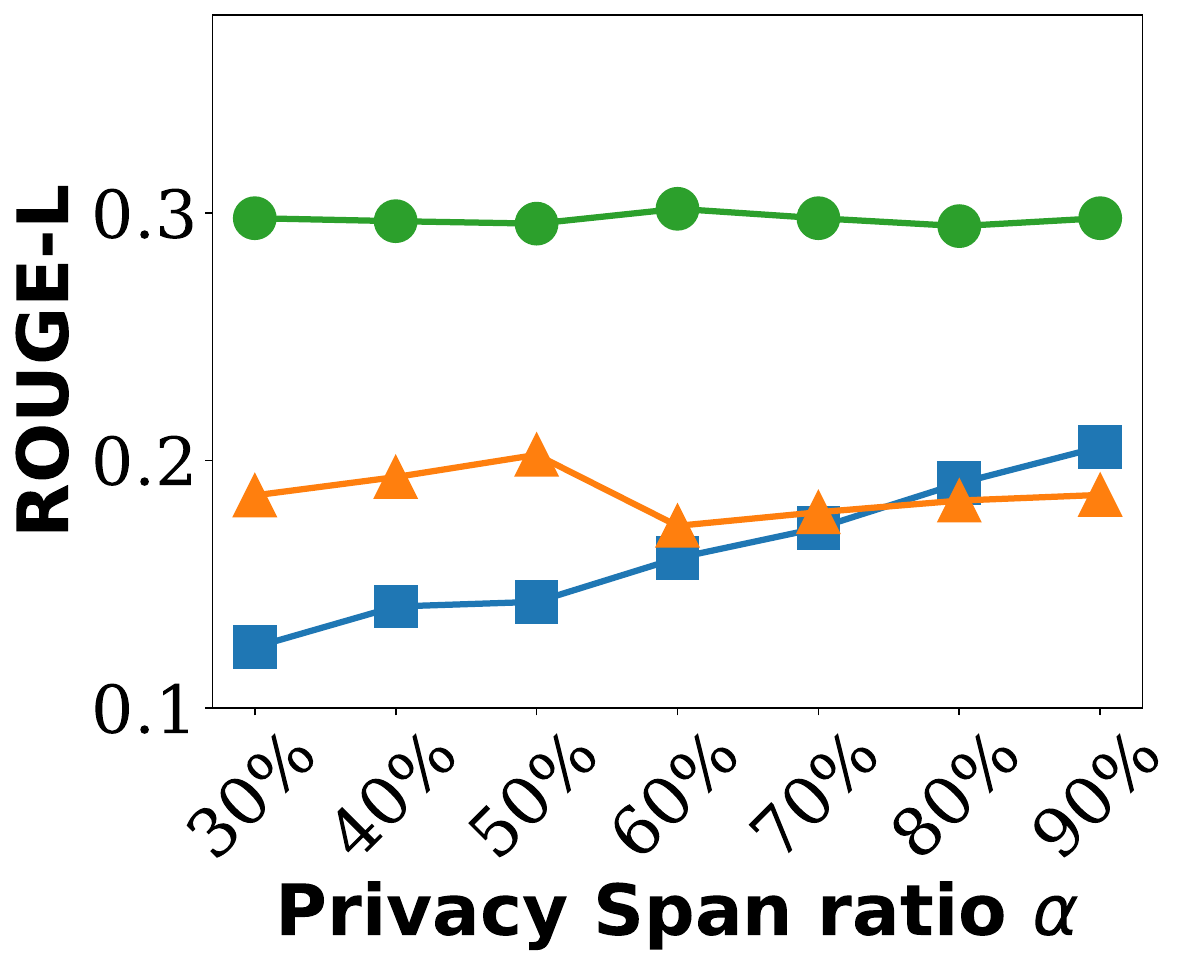}
        \label{fig:ea_pr}
    }
    \hfill
    \subfigure[Attribute Inference Attack]{
        \includegraphics[width=0.22\textwidth]{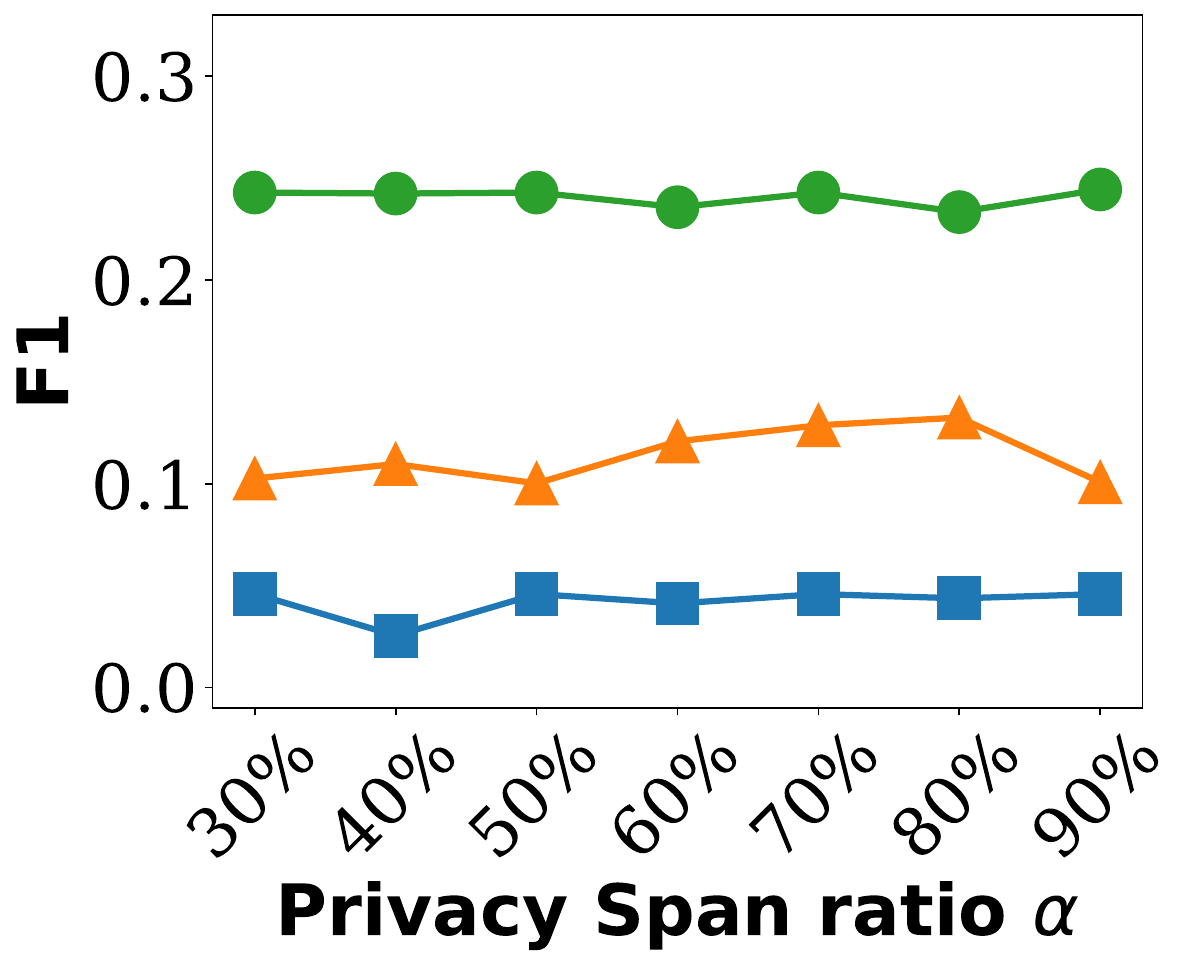}
        \label{fig:ai_pr}
    }
  \vspace{-2em}
  \vskip\baselineskip 
    \caption{(a) and (b) present the results of $d_\chi$-privacy method under the prompt injection attack and attribute inference attack under varying $d_\chi$-privacy percentages across three privacy-preserving datasets. 
    (c) and (d) show the results of PrivacyRestore for the embedding inverse attack and attribute inference attack under different privacy span ratios $\alpha$ on the same three datasets.}
    \label{fig:four_images}
\end{figure*}

\subsection{Ablation study}

\paragraph{Attention-Aware Weighted Aggregation.}
\label{app:EWAvsAWA}
\vspace{-0.1em}
To assess the effectiveness of the Attention-Aware Weighted Aggregation (AWA) component, we compare its performance and inference efficiency against Equal Weighted Aggregation (EWA). Unlike AWA, EWA generates the meta vector by summing all restoration vectors equally. 
As shown in Table \ref{tbl:no_weight}, EWA results in lower MC1, MC2, ROUGE-L, and LLM-J scores compared to AWA, indicating that equal weighting will diminish performance by amplifying irrelevant privacy spans.
\vspace{-1em}
\begin{table}[!htbp]
    \centering
    \def\arraystretch{1} 
    \resizebox{1\columnwidth}{!}{
    \scalebox{1}{
        \begin{tabular}{ l l |  ccccc  }
        \toprule
        \textbf{Datasets} & \textbf{Methods}  & \textbf{MC1 $\uparrow$} & \textbf{MC2 $\uparrow$} & \textbf{ROUGE-L $\uparrow$} & \textbf{LLM-J $\uparrow$} & \textbf{TP $\uparrow$} \\
        \cmidrule(lr){1-7} 
        \multirow{2}{*}{Pri-DDXPlus} & EWA & 53.84 & 51.12 & 26.32 & 4.29 & \textbf{26.35}    \\
        ~& \textbf{AWA}\cellcolor{lightgray!45} & \textbf{62.97}\cellcolor{lightgray!45} & \textbf{60.19}\cellcolor{lightgray!45} & \textbf{27.24}\cellcolor{lightgray!45} & \textbf{4.47}\cellcolor{lightgray!45} & 26.09\cellcolor{lightgray!45} \\

        \cmidrule(lr){2-7} 
        \multirow{2}{*}{Pri-NLICE} & EWA & 46.92 & 45.89 & 22.78 & 3.12 & \textbf{32.75} \\
         ~& \textbf{AWA}\cellcolor{lightgray!45} & \textbf{62.23} \cellcolor{lightgray!45}& \textbf{57.94}\cellcolor{lightgray!45} & \textbf{24.42}\cellcolor{lightgray!45} & \textbf{3.67}\cellcolor{lightgray!45} & 32.33\cellcolor{lightgray!45} \\

        \cmidrule(lr){2-7} 
        \multirow{2}{*}{Pri-SLJA} & EWA & 30.88 & 30.70 & 30.96 & 4.10 & \textbf{31.00} \\
         ~& \textbf{AWA}\cellcolor{lightgray!45} & \textbf{35.47}\cellcolor{lightgray!45} & \textbf{35.41}\cellcolor{lightgray!45} & \textbf{37.56}\cellcolor{lightgray!45} & \textbf{5.25}\cellcolor{lightgray!45} & 30.73\cellcolor{lightgray!45} \\
        \midrule
        \end{tabular}
    }
    }
    \vspace{-0.5em}
    \caption{
    Comparison of the performance and the inference efficiency between Equal Weighted Aggregation (\textbf{EWA}) and Attention-aware Weighted (\textbf{AWA}) Aggregation. 
    The best results are highlighted in \textbf{bold}.
    }
    \label{tbl:no_weight}
\end{table}

\vspace{-1.4em}
\paragraph{Other Ablation Studies.}
Furthermore, we evaluate PrivacyRestore’s performance with varying numbers of edited heads ($K$) and with an alternative LLM backbone (Llama-13b-chat) in Appendix \ref{app:abla}. 
The results in Table \ref{tbl:vary_heads} and Figure \ref{fig:13b} clearly demonstrate the effectiveness of our method.

\subsection{Extension Analysis of PrivacyRestore}
In this section, we analyze the extension PrivacyRestore to other more extreme scenarios.

\paragraph{Encountering Out-of-Set Privacy Spans.}
\label{sec:out_of_set}
Due to the long-tailed distribution of privacy spans shown in Appendix \ref{app:long_dis}, the core set covers most spans.
We further evaluated our method when encountering out-of-set spans. 
Specifically, we include only a subset of privacy span types in our core set. 
Table \ref{table:top_predefine} shows that our method still demonstrates superior performance, compared to No Restoration baseline.
More implementation details are shown in Appendix \ref{sec:new_privacy_spans}.

\paragraph{Users Unable to Determine Privacy Spans.}
\label{sec:varying_user_queries} 
Our method follows the principle of ``Information Self-Determination Rights'' allowing users to determine their own privacy spans. 
Even when users cannot or choose not to specify these spans, our method remains effective by integrating with existing \textbf{text sanitization techniques} \citep{Kan2023ProtectingUP,Chen2023HideAS}.
As shown in Table \ref{table:prires_sanitized}, our method can maintains superior performance, and details of implementation are provided in Appendix \ref{app:un_deter_privacy}.

\section{Conclusion}

We propose PrivacyRestore which protects the privacy within user inputs during inference in online LLM inference services. 
PrivacyRestore achieves privacy protection by directly removing privacy spans in the user input and then restoring information via activation steering.
PrivacyRestore provides a practical and efficient solution for protecting privacy while maintaining satisfactory performance and inference efficiency. 
We demonstrate that PrivacyRestore inherently addresses the linear growth problem of the privacy budget found in $d_\chi$-privacy. 
We curate three privacy-preserving datasets covering medical and legal fields, and PrivacyRestore achieves strong performance and inference efficiency across all datasets. 
Additionally, we implemented various attack methods, and the attack results demonstrate PrivacyRestore's robust privacy protection capabilities.

\newpage
\section*{Limitations}
This section aims to highlight the limitations of our work and provide further insights into the research in this area. 

One limitation is that we only evaluate our method in the medical and legal domains and additional domains could be explored to validate its effectiveness.

Another limitation is that more attack methods could be explored to assess the privacy protection of our approach. 
While we have implemented most of the current advanced attack methods, to the best of our knowledge, there may be others yet to be tested. 
Additionally, more advanced attack methodologies may emerge in the future, which will also need to be evaluated.

\section*{Ethics Statement}
We adhere to the ACL Ethics Policy and have conducted our research using publicly available repositories and datasets. In the PrivacyRestore framework, we have adhered to rigorous ethical standards to safeguard user privacy and uphold data security. All three datasets (Pri-SLJA, Pri-NLICE and Pri-DDXPlus) utilized in this research are sourced exclusively from publicly available repositories, ensuring that these datasets are devoid of any personally identifiable information and minimizing potential privacy risks. Our methodology does not access or reconstruct the original identifiable data or its sources. This ensures that the research does not infringe upon individual privacy rights. 

However, due to the fact that we employed multiple LLMs in this study, such as ChatGPT, Qwen and GPT-4. The findings may be influenced by the inherent assertiveness, linguistic patterns, and diverse biases characteristic of these LLMs.

\section*{Acknowledgment}
This research was supported by the National Natural Science Foundation of China (62406114,62306117,62472181), the National KeyR\&D Project fromMinister of Science and Technology (2024YFA1211500), 
the Fundamental Research Funds for the Central Universities (2024ZYGXZR074),  
the Guangdong Basic and Applied Basic Research Foudation (2025A1515011413,2024A1515010220), 
the Guangzhou Basic and Applied Basic Research Foundation (2024A04J3681), 
the GJYC program of Guangzhou (2024D03J0005) 
and the South China University of Technology-TCL Technology Innovation Fund.

The work described in this paper was conducted in full or in part by Dr. Haoran Li, JC STEM Early Career Research Fellow, supported by The Hong Kong Jockey Club Charities Trust.
\bibliography{acl}

\clearpage
\newpage
\appendix
\section*{Appendix Overview}
The appendix is divided into two parts: Appendix~A--I provide backup theoretical explanations of PrivacyRestore, while Appendix~J--X present additional experimental results on PrivacyRestore from different aspects.

\section{Notations}
\label{sec:notations}
Here we present all notations used in our paper in Table \ref{tab:symbols}.

\begin{table*}[htpb]
    \centering
    \resizebox{1\textwidth}{!}{
    \begin{tabular}{l|l}
        \toprule
        \textbf{Notations} & \textbf{Definitions} \\
        \midrule
        $c$ & A privacy span type. \\ 
        $\mathcal{C}$ & All possible privacy span types. \\
        $s$ & A single privacy span. \\
        $\mathcal{S}_q$ & All privacy spans in user query $q$. \\
        $h$ & A single edited head. \\
        $\mathcal{H}_k$ & The common top-K heads set. \\
        $\mathcal{H}_a$ & The set of all heads. \\
        $\mathcal{H}_k^c$ & The top-K heads set of the privacy span type $c$. \\
        $L_h$ &  The score list of the head $h$ across all privacy spans. \\ 
        $K$ & The number of selected edited heads. \\
        
        $\mathcal{F}^{c}_{h}$ & The probe of privacy span type $c$ on head $h$. \\
        $\theta^{c}_{h}$ & The parameters of the probe $\mathcal{F}^{c}_{h}$. \\
        $\textbf{u}_h$ & The output hidden state on head $h$. \\
        $\bar{\textbf{u}}_h$ & The output hidden state after restoration on head $h$. \\

        $r_h^c$ & The restoration vector for privacy span type $c$ on head $h$. \\
        $\Theta$ & All restoration vectors for all privacy spans on all edited heads. \\

        $\lambda$ & The tradeoff hyperparameter of ORPO loss. \\ 
        $w_s$ & The weight of privacy span $s$. \\
        $n$ & The number of tokens in the user query. \\
        $n_h$ & The number of heads in the lightweight model. \\
        $\text{Attn}_h(x, y)$ & The attention score of $y$ attending to $x$ on head $h$. \\
        $Z_h$, $Z_h^\prime$ & Any two normalized weighted sums of restoration vectors on head $h$. \\ 
        $\mathcal{R}$ & The meta vector. \\
        $\mathcal{R}_h$ & The part of the meta vector for head $h$. \\
        $\mathcal{N}$ & The added noise on the normalized weighted sums for meta vector construction. \\
        $I$ & The user inputs in the training set. \\
        $I_{all}= \{I_1, ..., I_m\}$ & All user inputs in the training set. \\
        $Y_c=\{y_1, ..., y_m\} $ & The labels indicating whether the corresponding input contains the privacy span of type $c$. \\
        $m$ & The size of training set. \\
        $I$, $I^\prime$ & Any two user inputs. \\
        $\{i_1, ..., i_n\}$ & The tokens of the input $I$. \\
        $\{e_1, ..., e_n\}$ & Corresponding token embeddings of the input $I$. \\
        $O = \{o_1, ..., o_n\}$ & The possible output sets for $I$, with each one representing a single output.\\
        $\hat{I}$ & The incomplete user input with all privacy spans removed in the training set. \\
        $\hat{I}_{all}=\{\hat{I}_1, ..., \hat{I}_m\}$ & All user inputs with privacy spans removed in the training set. \\
        $a$ & The output given the complete input $I$. \\
        $\hat{a}$ & The output given the incomplete input $\hat{I}$. \\

        $q$ & The user query during inference. \\
        $\hat{q}$ & The incomplete user query with all privacy spans removed during inference. \\


        $\epsilon$ & The privacy hyperparameter. \\
        $\tau$ & The generation temperature. \\
        $\delta$ & The privacy hyperparameter. \\

        $n_{ps}$ & The number of tokens associated with the privacy spans in the user query. \\
        $\alpha$ & The proportion of privacy spans selected for protection. \\
        $d_\chi$ & Any distance function used by $d_\chi$-privacy. \\

        $d_e$ & The distance between token embeddings. \\
        $d_z$ & The distance between normalized weighted sums. \\

        \toprule

    \end{tabular}
    }
    \caption{Definitions of all notations used in our paper.}
    \label{tab:symbols}
    
\end{table*}



\section{Related Works}


In this section, we introduce the related works on user input protection methods, which are currently divided into two categories: SMPC-based methods and DP-based methods.


Here, we provide a more detailed introduction to Secure Multi-Party Computation (SMPC) and Differential Privacy (DP).

\label{app:appendix_related_work}
















\subsection{Secure Multi-Party Computation (SMPC)}

Secure multi-party computation (SMPC) methods utilize multi-party encryption algorithms to enable collaborative computation among multiple parties while protecting the privacy of their data.

However, most nonlinear operations in LLMs cannot directly support secure multi-party computation.

To address this challenge, current SMPC methods focus on two optimization directions: model structure-oriented optimization and protocol-oriented optimization.

The model structure-oriented approach aims to replace SMPC-unfriendly nonlinear operations with SMPC-friendly alternatives.

For instance, MPC-Former \citep{li2023mpcformer} approximates nonlinear operations in Transformer using polynomials and maintains performance through model distillation.

MERGER \citep{liang2024merge} integrates previous techniques to natural language generation (NLG) tasks by bypassing embedded computation and reorganizing linear operations in Transformer modules, further enhancing computational efficiency and model performance.

In contrast, the protocol-oriented approach focuses on designing efficient SMPC operators for nonlinear operations in LLMs while preserving the original model structure.

Recent works \cite{mao2022iron, liu2023encrypt, zheng2023primer, kanav2023sigma} have improved the efficiency of nonlinear operations in privacy-preserving LLMs inference by utilizing various SMPC protocols, such as confusion circuit and function secret sharing.

Although SMPC-based methods can be applied to protect user inputs during model inference, they still suffer from large inference time overhead.


For example, inference on the RoBERTa-Base model takes 168.43 seconds \citep{hao2022iron}, making current SMPC methods impractical for online LLM inference services.

\subsection{Differential Privacy (DP)}

Differential Privacy (DP), as introduced by \citet{cyn2016cali}, is designed to protect individual privacy by preventing attackers from identifying specific participants in a dataset. 

Several variants of DP have been developed to enhance privacy protection across various settings, adapting the core principles of DP to different types of data and threat models.

Notable examples include Centralized Differential Privacy (CDP), Local Differential Privacy (LDP), and $d_{\chi}$-privacy.

CDP \citep{cyn2016cali} operates under the assumption that all data has been stored in a central repository.

It guarantees that attackers cannot distinguish between any two adjacent repositories based on query results.

In contrast, LDP \citep{john2013local} provides a stronger guarantee, ensuring that attackers cannot distinguish between any two adjacent inputs. \citet{justus2022limit} and \citet{sai2023ldp} propose using paraphrasing techniques to achieve LDP on user inputs.

$d_{\chi}$-privacy \citep{feyisetan2019privacy}, a relaxed version of LDP, incorporates metrics that measure the similarity between inputs, allowing for more flexible control over the privacy budget. The formal definition of $d_{\chi}$-privacy Mechanism is provided in Appendix \ref{app:method_preliminary_dx_privacy}.

As proposed by \citet{justus2022limit}, applying $d_\chi$-privacy to all tokens in user inputs, known as word-level privatization, suffers from the linear growth problem of the privacy budget.

This means that as the length of the protected text increases, the privacy protection performance of $d_\chi$-privacy decreases.

\section{Preliminaries for Methodology}
\label{app:method_preliminary}
The $d_{\chi}$-privacy mechanism and activation steering technique are two crucial components of our method. Here, we provide a more detailed illustration of these techniques for a better understanding of our method.
\subsection{$d_{\chi}$-privacy Mechanism}
\label{app:method_preliminary_dx_privacy}
$d_{\chi}$-privacy mechanism \citep{feyisetan2019privacy} is a variant of the differential privacy mechanism designed to protect the privacy and incorporate a distance measure into the privacy budget. 
The detailed definition of $d_\chi$-privacy is as follows:
\begin{definition}
\textbf{(}$\bm{d_{\chi}}$-\textbf{privacy mechanism).}
\textit{A randomized mechanism $\mathcal{M} : \mathcal{I}\to\mathcal{O}$ fulfills $\epsilon$-$d_{\chi}$-privacy if for all adjacent inputs $I, I^{\prime} \in \mathcal{I}$ and all possible outputs $O \subset \mathcal{O}$, the following condition holds:}
\begin{eqnarray*}
\label{def:dx_privacy}
    \mathbb{P}\left(\mathcal{M}(I)\in O\right)
    \leq\exp(\epsilon d_{\chi}(I,I^{\prime}))
    \mathbb{P}\left(\mathcal{M}(I^{\prime})\in O\right),
\end{eqnarray*}
where $d_{\chi}$ is a distance function defined on $\mathcal{I}$.
\end{definition}
Numerous prior works have applied the $d_\chi$-privacy mechanism \citep{konstantinos2013broad, pazii2018local} to word embeddings to achieve word-level privatization \citep{feyisetan2020privacy, feyisetan2019privacy, xu2020diff, bo2021er}. 
In our approach, we employ the $d_\chi$-privacy mechanism to protect the meta vector, preventing privacy leakage from the meta vector.

To implement the $d_{\chi}$-privacy mechanism on the meta vector/token embeddings, noise must typically be added to it, as shown below:
\begin{eqnarray}
    \mathcal{R} &=& Z + \mathcal{N}, \\
    \mathbb{P}(\mathcal{N}) &\propto& \exp(-\epsilon||\mathcal{N}||),
\end{eqnarray}
where $Z$ is the unprotected meta vector/token embeddings, $\mathcal{N}$ is the added noise, $\mathcal{R}$ is the protected meta vector/token embeddings and $\epsilon$ is the privacy parameter of the mechanism.
According to \citet{feyisetan2019privacy}, to sample the noise $\mathcal{N}$ from its distribution, we can compute it as follows:
\begin{eqnarray}
   \textbf{v} &\in& \{v\in\mathbb{R}^n:||v||=1\} \\
    \mathbb{P}(\textbf{l}) &\propto& \frac{\textbf{l}^{n-1}e^{-\epsilon \textbf{l}}}{\Gamma(n)\epsilon^{-n}}, \\
    \mathcal{N} &=& \textbf{l} \cdot \textbf{v},
\end{eqnarray}
where $n$ is the size of the meta vector and $\epsilon$ is the privacy parameter.

\subsection{Activation Steering Technique}
Activation steering methods \citep{li2023iti, turner2023activation, hernandez2023inspecting} control the behavior of LLM by modifying their activations during the inference stage.
It serves as a crucial part of our methodology to restore information contained within the removed privacy spans during LLM inference. 
Typically, the attention mechanism \citep{vaswani2017attention} in LLM is responsible for capturing contextual information, and it can be expressed as:
\begin{eqnarray}
    q &=& W_q \cdot \textbf{i}, \\
    \textbf{u} &=& \text{Softmax}(\frac{q \cdot K^T}{\sqrt{d_k}}) \cdot V,
\end{eqnarray}
where $\textbf{i}$ is the input hidden state, $\textbf{u}$ is the output hidden state, $W_q$ is the query weight matrix, $K$ is the key of the context and $V$ is the value of the context and $d_k$ is the dimension of the key. Activation steering methods add some steering vectors into the output hidden state and, in our methods, we add the meta vector into the output hidden state to restore information, which can be expressed as:
\begin{equation}
    \textbf{u} = \textbf{u} + \mathcal{R},
\end{equation}
where $\mathcal{R}$ is the meta vector.

\section{Selecting the Most Relevant Heads}
\label{app:probe_techniques}
In this section, we provide the implementation details of the probe technique \citep{alain2016understanding, tenney2019bert, belinkov2022probing} to identify the most relevant attention heads for each type of privacy span.

Let $I_{all} = {I_1, ..., I_m}$ represent the user inputs in the training set, where $m$ is the size of the training set. For a given privacy span type $c$, let $Y_c = {y_1, ..., y_m}$ represent the corresponding labels, where $y_i = 1$ if and only if the input $I_i$ contains a privacy span of type $c$.

For each user input $I_i$, we record the hidden state of the last token on each attention head. We then train a binary classifier for each head, tailored to the privacy span type $c$, as the probe. The probe takes the hidden state of the last token as input and predicts whether the input contains the privacy span of type $c$.
The probe is formulated as: 
\begin{equation} 
\mathcal{F}^{c}_{h} (\textbf{u}_{h}) = \sigma (\theta_{h}^{c} \cdot \textbf{u}_{h}),
\end{equation} 
where $\mathcal{F}^{c}_{h}(\cdot)$ is the probe of privacy span type $c$ on head $h$, $\textbf{u}_{h}$ is the hidden state of the last token on head $h$, $\theta_{h}^{c}$ are the parameters of the probe, and $\sigma(\cdot)$ indicates the sigmoid function.

A probe $\mathcal{F}^{c}_{h}(\cdot)$ with higher accuracy indicates a stronger correlation between the head $h$ and the privacy span type $c$. 
Therefore, we only select the top $K$ attention heads, with the highest accuracies, as the most relevant heads for the privacy span type $c$.

\section{The Algorithm of Common Top-K Selector}
\label{sec:selector}
In this section, we present the detailed implementation of Common Top-K Selector algorithm, as shown in Algorithm \ref{alg:agg_head}.

Firstly, we initialize an empty score list $L_h$ for each head. 
Secondly, each privacy span type $c$ has its corresponding top-K heads set $\mathcal{H}_k^c$. 
For each head $h$ in $\mathcal{H}_k^c$, we append $\text{Score}(h, \mathcal{H}_k^c)$ into its score list $L_h$.
$\text{Score}(h, \mathcal{H}_k^c)$  is defined as the rank of head $h$ among $\mathcal{H}_k^c$ in ascending order based on the accuracy of the probe associated with the head $h$ and privacy span type $c$.
Thirdly, we calculate the average value of each score list $L_h$ as the score of the corresponding head $h$.
Finally, we sort all heads in the LLM by the scores and pick up top-K heads as the common top-K head set $\mathcal{H}_k$. 

\begingroup
\setlength{\topsep}{0.1px}    
\setlength{\partopsep}{0pt} 
\begin{algorithm}[htbp] 
    \caption{Common Top-K Selector}
    \label{alg:agg_head}
        \begin{flushleft}
            \textbf{Input:} $\mathcal{S}$ is the set of privacy spans; $\mathcal{H}_a$ is the set of all heads; 
            $\mathcal{H}_k^c$ denotes the set of top-K heads corresponding to the privacy span type $c$; $\text{Score}(h, \mathcal{H}_{k}^c)$ return the rank of head $h$ among $\mathcal{H}_k^c$ in ascending order based on the accuracy of the probe associated with the head $h$ and privacy span type $c$. The score of the head with lowest accuracy is $1$. The score of the head with highest accuracy is $K$.
        \end{flushleft}
            \begin{algorithmic}[1]
                \STATE Initialize an empty score list $L_h = [\;]$ for each head $h$ in $\mathcal{H}_a$. \\
                \FOR{ $c\;$ \text{in} $\;\mathcal{C}$} 
                    \FOR{$h\;$ \text{in} $\;\mathcal{H}_k^c$}
                    \STATE Append $\text{Score}(h, \mathcal{H}_k^c)$ into $L_h$. 
                    \ENDFOR
                \ENDFOR
                 \FOR{$h\;$ \text{in} $\;\mathcal{H}_a$}
                   \item $\text{score}_h = \text{average}(L_h)$
                \ENDFOR
                \STATE Sort $\mathcal{H}_a$ according to $\text{score}_h$ and select top $K$ heads to obtain \textbf{common top-K head set} $\mathcal{H}_k$.
            \end{algorithmic}
        \begin{flushleft}
            \textbf{Output:} $\mathcal{H}_k$ is the common top-K head set.
        \end{flushleft}
\end{algorithm}   
\endgroup

\section{Details of the Training Process}
\label{sec:details_train}
In our method, we use the ORPO loss \cite{orpo} to train the restoration vectors, which are employed to restore information in the removed privacy spans. 
The training objective is to ensure that, despite receiving incomplete inputs with all privacy spans removed, the model can still generate high-quality outputs similar to those produced from intact inputs by utilizing these restoration vectors.

Assuming that $\Theta$ is the trainable restoration vectors, $\hat{I}$ denotes the input with privacy spans removed, $\hat{I}_{all} = \{\hat{I}_{1},\cdots,\hat{I}_{m}\}$ represents the training set of incomplete inputs, $a$ is the initial output give the complete input, $\hat{a}$ is the output given the incomplete input with privacy spans removed, then the training loss of our method can be express as:


\begin{equation}
\begin{aligned}
    \text{ratio}(a|\hat{I}; \Theta) &= \frac{\mathbb{P}(a|\hat{I}; \Theta)}{1-\mathbb{P}(a|\hat{I}; \Theta)},\\
    \mathcal{L}_{\text{ORPO}} 
    &= \sum_{\hat{I} \in \hat{I}_{all}} - \log \mathbb{P}(a|\hat{I}; \Theta)  \\
    & - \lambda \log \sigma \left( \log \frac{\text{ratio}(a|\hat{I}; \Theta)}{\text{ratio}(\hat{a}|\hat{I}; \Theta)} \right),
\label{eq:loss}
\end{aligned}
\end{equation}
where $\lambda$ is the hyperparameter that controls the weight of the loss term and the  $\mathbb{P}(a|\hat{I}; \Theta)$ is the probability of the model generating the initial output $a$ given the intact input after being restored by $\Theta$, and $\mathbb{P}(\hat{a}|\hat{I}; \Theta)$ is the probability of generating $\hat{a}$. 
After training the restoration vectors using the above loss, these vectors can effectively restore the information in the missing privacy spans and guide the model to generate outputs similar to those produced from intact inputs, \textbf{even though no privacy spans are present in the input indeed}.

We also provide a training example for better understanding of our training process.
As shown in the loss function in Equation \ref{eq:loss}, each training sample will contain the incomplete input without privacy spans $\hat{I}$, the intact input $I$, the output $a$ given intact input $I$ and the output given the incomplete input $\hat{I}$. An example of a training sample is presented in Figure \ref{fig:training_samples}.
\begin{figure}[h]
  \centering
  \raggedleft
    \begin{minipage}{0.98\columnwidth}
    \includegraphics[width=1\textwidth]{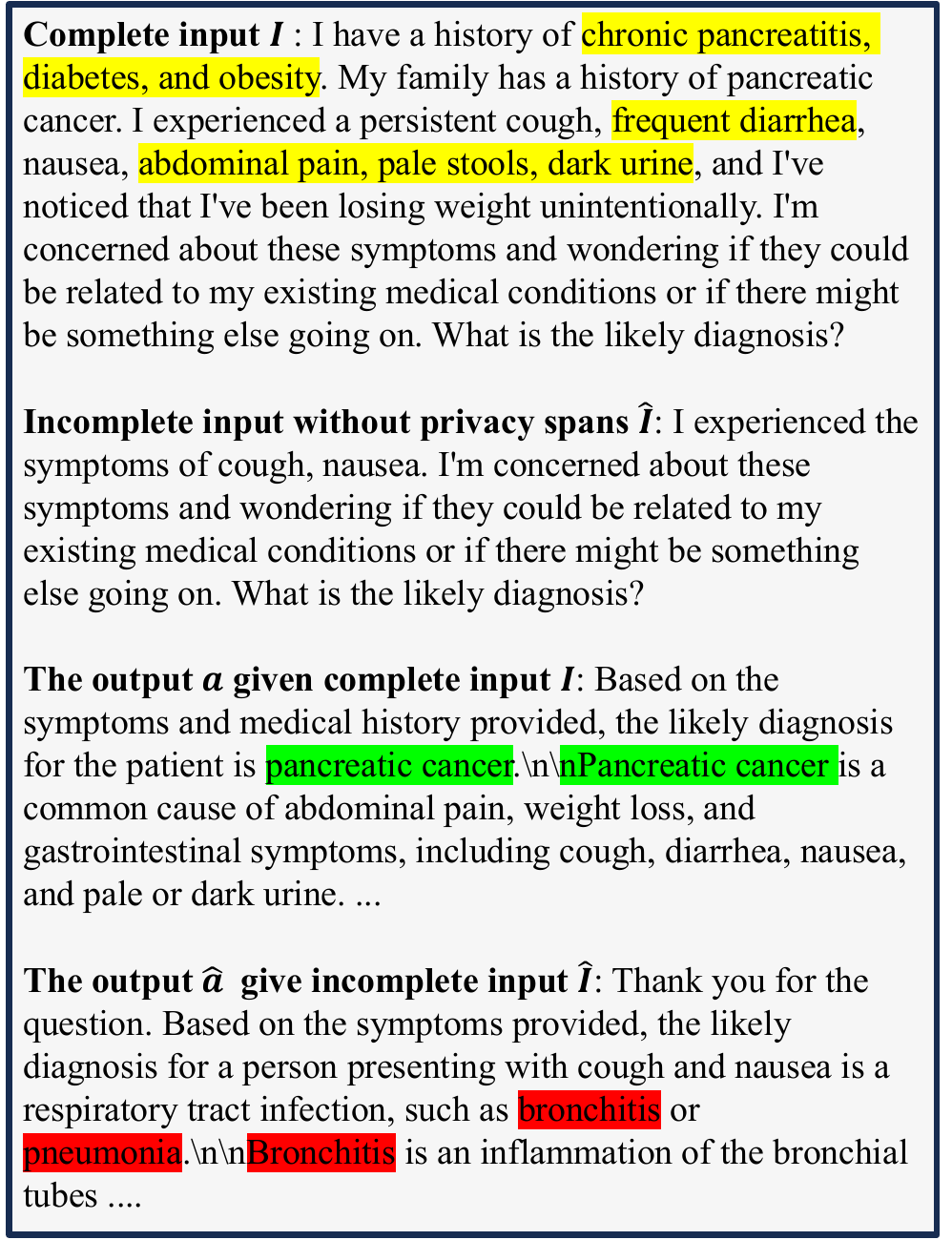}
    \label{fig:training_samples_legend}
  \end{minipage}
  \vspace{-1.5em}
  \caption{A training sample in our framework. Text highlighted with a yellow background represents the privacy spans in user inputs. Text highlighted with a green background indicates the correct diagnosis. Text highlighted with a red background denotes the incorrect diagnosis. }
  \label{fig:training_samples}
\end{figure}

\section{Usage of the Lightweight Model on the Client Side}
As shown in Section \ref{sec:inference}, our method utilizes the lightweight model (e.g. Bert-based-uncased \citep{devlin2019bert}) to classify the privacy spans in the user query and compute the importance score of these privacy spans for conducting the following Attention-aware Aggregation (AWA).
The detailed implementations are as follows:

\subsection{For Classifying Privacy Span Types}
\label{app:usage_1}
Each privacy span type can be expressed in various forms within the user query, for example,``\textit{fever}'' may be represented as``\textit{elevated body temperature}''. 
After the user identifies the privacy spans in the query, we should classify these spans into those predefined types from the set $\mathcal{C}$.

Firstly, we use a lightweight Bert-based-uncased model on the client side to first extract the vector representation of the privacy span. 
Specifically, we compute the mean of the hidden states from the last layer across all tokens within the privacy span to obtain the vector representation.
We construct a multi-layer perceptron (MLP) classifier, consisting of an input layer, two hidden layers, and an output layer.
The MLP classifier takes the vector representation of the privacy span as input, and the output label corresponds to the privacy span type.
During the training process, we will fix the Bert-based uncased model while only training the MLP classifier.

\subsection{For Computing the Importance Score of Privacy Spans}
\label{app:usage_2}
Each privacy span in the user query should have a distinct importance weight and we also utilized the Bert-base-uncased model to assess the importance weights for the privacy spans.
To be specific, we compute the average received attention of privacy span $s$ across all attention heads and all tokens in the user query as the importance score $w_s$.
Assume $s$ is the privacy span $s$, $q$ is the user query, and then the importance weight of the privacy span $w_s$ is calculated as:
\begin{eqnarray}
    \label{eq:weight}
    w_s &=& \frac{1}{n} \frac{1}{n_h} \sum_{t=1}^{n}\sum_{h=1}^{n_h}\text{Attn}_{h}(s, q_t),
\end{eqnarray}
where $n$ is the number of tokens in the query, $n_h$ is the number of attention heads in the lightweight model, $q_t$ is the $t$-th token of $q$, and $\text{Attn}_h(s, q_t)$ denotes the attention score of $q_t$ attending to the privacy span $s$. 
Higher $w_s$ indicates that privacy span $s$ receives more attention from other tokens in the user query $q$, reflecting greater importance.


\section{Proof of Theorem~\ref{theo_2}}
\label{sec:fur_theo_2}
As shown in Figure \ref{fig:algorithm}, during the inference stage, only the meta vector and the incomplete query with privacy spans removed are transmitted from the client to the server.
The incomplete query does not contain any privacy-sensitive information and is secure for the user.
The meta vector contains information about all privacy spans and could be vulnerable to adversaries attempting to reverse-engineer these spans, requiring privacy protection.

PrivacyRestore protects the meta vector by adding noise $\mathcal{N}$ which is sampling from the distribution $p(\mathcal{N}) \propto \exp(-\epsilon\|\mathcal{N}\|)$, before transmission, as shown in Eq \ref{eq:add_noise}.
Next, we will demonstrate that \textbf{injecting noise in this manner adheres to the definition of $d_\chi$-privacy} and effectively protects the user privacy contained in the meta vector.

Assume $Z$ represents the meta vector before adding noise, $\mathcal{R}$ denotes the meta vector after adding noise, as shown in Eq \ref{eq:orgianl_meta_vec} and \ref{eq:add_noise}. 
The process of adding noise can be represented by $\mathcal{M}$. 
Then, the possibility that $Z$ becomes $\mathcal{R}$ after adding noise $\mathcal{N}$ is 
\begin{equation}
\begin{aligned}
    \mathbb{P}(\mathcal{M}(Z) = \mathcal{R}) &= \mathbb{P}(Z + \mathcal{N} = \mathcal{R}) \\ 
    &= \mathbb{P}(\mathcal{N} = \mathcal{R} - Z) \\ 
    &= \exp(-\epsilon || \mathcal{R} - Z ||).
\end{aligned}
\end{equation}
Then for any two meta vectors before adding noise, $Z$ and $Z^\prime$, we have:
\begin{equation}
\begin{aligned}
    \frac{\mathbb{P}[\mathcal{M}(Z) = \mathcal{R}]}{\mathbb{P}[\mathcal{M}(Z^\prime) = \mathcal{R}]} &=
    \frac{\exp(-\epsilon || \mathcal{R} - Z ||)}{\exp(-\epsilon || \mathcal{R} - Z^\prime ||)} \\ 
    &= \exp(\epsilon (|| \mathcal{R} - Z^\prime ||
    \\& \qquad \qquad - || \mathcal{R} - Z ||) ) \\ 
    &\leq \exp(\epsilon ||Z^\prime - Z||).
\end{aligned}
\end{equation}
According to the definition of $d_\chi$-privacy in Appendix \ref{app:method_preliminary_dx_privacy}, the mechanism $\mathcal{M}$ satisfies $d_\chi$-privacy. 
In other words, by adding noise $\mathcal{N}$, adversaries cannot infer the initial meta vector $Z$ from the meta vector after adding noise $\mathcal{R}$, even if $\mathcal{R}$ is intercepted. 
Moreover, the privacy budget of our methods is $\epsilon ||Z^\prime - Z||$. And considering that $Z$ is the normalization of the weighted sum of restoration vectors, as shown in Eq. \ref{eq:orgianl_meta_vec}, then we have:
\begin{equation}
\begin{aligned}
    \frac{\mathbb{P}[\mathcal{M}(Z) = \mathcal{R}]}{\mathbb{P}[\mathcal{M}(Z^\prime) = \mathcal{R}]} &\leq \exp(\epsilon ||Z^\prime - Z||) \\
    &\leq \exp(2\epsilon)
\end{aligned}
\end{equation}
Thus, the privacy budget of our method is $2\epsilon$, independent of the input length $n$ 
and solely depends on the hyperparameter $\epsilon$.
In summary, PrivacyRestore fulfills $d_\chi$-privacy and provides a privacy budget $2\epsilon$ which is independent of the input length and inherently addresses the problem of the linear growth of privacy budget.

\section{Brief Proof of Output Protection}
\label{app:proof_output_pro}

It has been proved by Appendix A of \citet{Utpala2023LocallyDP} and Section 4.2 of \citet{Mattern2022TheLO} that sampling-based generation can prevent the privacy leakage via the generated output via the Exponential Mechanism.
Here, we provide a brief proof that sampling-based generation adheres to the Exponential Mechanism \citep{Mc2007Mecha}, ensuring security for the generated output.

Assume that $Q$ is the user query,
$\mathcal{V}$ is the whole token vocabulary, $u \in \mathbb{R}^{|\mathcal{V}|}$ is the output logit, $u_t$ is the logit for the token $t$ in $\mathcal{V}$ and $\mathcal{M}$ denotes the sampling based generation.
Recall that, during sampling-based generation, the logit $u$ should be processed by the softmax layer and then be sampled to obtain the output. 
If $T$ is the sampling temperature and $Pr[\mathcal{M}(Q)=t]$ indicates the probability of generating the token $t$, then the softmax layer can be expressed by:
\begin{equation}
\begin{aligned}
\label{eq:softmax_dp}
Pr[\mathcal{M}(Q)=t] = \frac{\exp(u_t/T)}{\sum_{j=1}^{|\mathcal{V}|} \exp(u_j/T)}
\end{aligned}
\end{equation}
Let recall the Exponential Mechanism \citep{Mc2007Mecha}, assuming $u$ is the utility function and $\Delta u$ is the sensitivity of $u$, then $\mathcal{M}$ satisfy the Exponential Mechanism if and only if
\begin{equation}
\begin{aligned}
\label{eq:exponential_mechanism}    
Pr[\mathcal{M}(Q)=t] 
&= \frac{\exp(\epsilon u(Q,t)/2\Delta u)}{\sum_{j=1}^{|\mathcal{V}|} \exp(\epsilon u(Q,j)/2\Delta u)} 
\\&
\propto \exp(\epsilon u(Q,t)/2\Delta u)
\end{aligned}
\end{equation}
By Comparing \ref{eq:softmax_dp} and \ref{eq:exponential_mechanism}, we can find that \textbf{the sampling from softmax layer follows the definition of Exponential Mechanism}, where $u(Q,t)$ and $u_t$ are different expressions of the same thing. 
Furthermore, according to the fact that the privacy budget of Exponential Mechanism is $\epsilon$, we can conclude that \textbf{the privacy budget of sampling-based generation is $2\Delta u/T$}. The privacy budget decreases with the increasing temperature, indicating that higher temperatures will bring better privacy protection.

\section{Datasets}
\label{app:dataset}

Based on the existing benchmarks, such as DDXPlus \citep{ars2022ddxplus} and NLICE \citep{al2023nlice} for medical diagnosis, and SLJA \citep{wen2023slja} for legal judgment, we construct three privacy-preserving datasets, Pri-DDXPlus, Pri-NLICE and Pri-SLJA, to evaluate the performance of various privacy-preserving methods.
In this section, we will introduce the detailed construction process of these three privacy-preserving datasets and provide some statistical information about them.

\subsection{Construction Process}
The total construction process of these privacy-preserving datasets consists of four stages: \textbf{Extraction of Privacy Spans}, \textbf{Rewriting Queries for Diversity}, \textbf{Assigning Options} and \textbf{Filtering Dataset}.
The details of these four stage are as follows:
\label{app:construction}

\paragraph{Extraction of Privacy Spans:}
We used GPT-4 \citep{achiam2023gpt} to classify symptoms in DDXPlus and NLICE, as well as case details in SLJA, into five levels ranging from non-sensitive to highly sensitive. 
The assessment prompt template is shown in Appendix \ref{sec:class_privacy}.
A higher level indicates that the symptom or case detail is more sensitive. 
We define all symptoms and case details with a sensitivity level greater than 3 as privacy spans.

\paragraph{Rewriting Queries for Diversity:}
The symptom descriptions and case details in the original DDXPlus, NLICE, and SLJA datasets are highly fixed. However, in real-world scenarios, these descriptions are typically more diverse. 
To address this gap, we utilized GPT-4 \citep{achiam2023gpt} to rewrite the user queries in these datasets, ensuring more varied descriptions and differing question formats while preserving the original meaning of the queries.
The rewrite prompt template is provided in Appendix \ref{sec:rewrite_queries}.
Here, we provide a rewrite example that demonstrates how rewriting the user query significantly increases the diversity of query descriptions, as shown in Figure \ref{fig:varying_format}.

\begin{figure}[h]
  \centering
  \raggedleft
    \begin{minipage}{0.98\columnwidth}
    \includegraphics[width=1\textwidth]{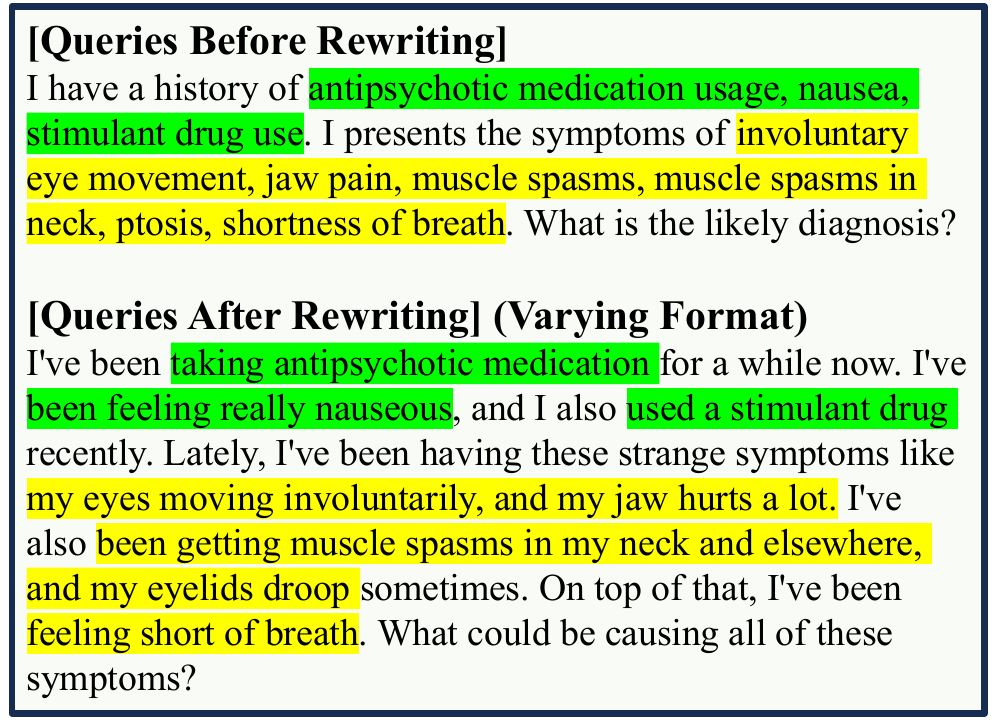}
    \label{fig:varying_format_legend}
  \end{minipage}
  \vspace{-1.5em}
  \caption{A rewrite example displays the diversity enhancement in medical queries. Text highlighted with green background indicates medical history, while yellow background denotes symptoms.}
  \label{fig:varying_format}
\end{figure}

\paragraph{Assigning Options:} 
To evaluate the performance of different privacy-preserving methods, we assign each sample a correct answer along with three randomly selected incorrect options. 
For DDXPlus and NLICE, we randomly select three diagnosis results to combine with the correct diagnosis as the choices. 
In the SLJA dataset, we randomly select three legal judgments to pair with the correct one as the options.

\paragraph{Filtering Dataset:} 
The initial dataset is extensive, and we observed that for most samples, removing all privacy spans often yields outputs similar to those obtained when privacy spans are provided.
Privacy preserving for these samples is meaningless because users can directly hide those privacy spans and obtain approximate result outputs.
In real-world scenarios, sensitive privacy spans often play a crucial role in medical diagnoses and legal judgments, making privacy preservation highly valuable.
Our dataset is designed to benchmark various privacy-preserving methods and must include samples where privacy spans are crucial for generating outputs.
We utilize the KL divergences to measure the importance scores of samples.
We calculate the KL divergence between the model output distributions with and without the privacy symptoms included.
A higher KL divergence indicates that the absence of sensitive privacy spans may lead to different or incorrect outputs. 
We selected only samples with high KL divergence to construct the privacy-preserving datasets.
As a result, we curated three privacy-preserving datasets: Pri-DDXPlus and Pri-NLICE for medical diagnosis, and Pri-SLJA for legal judgment.

\subsection{Statistical Information}
\label{sec:dataset}

We show the statistics of the obtained Pri-DDXPlus, Pri-NLICE and Pri-SLJA datasets in Table \ref{tbl:statistics}. 
We tally the number of user queries, privacy span types, and privacy spans count.
In Pri-DDXPlus and Pri-NLICE, the privacy spans are the symptoms, and the answers are the diagnoses.
In Pri-SLJA, the privacy spans are the case details, and the answers are the legal judgments.

Pri-DDXPlus commonly contains more sample instances and more privacy span types compared to Pri-NLICE and Pri-SLJA.
\begin{table*}[t]
    \centering
    \def\arraystretch{1} 
    \resizebox{0.75\textwidth}{!}{ 
        \large
        \begin{tabular}[h]{ll|cccc}
            \toprule 
            \textbf{Datasets} & \textbf{Dataset Split} & \textbf{User inputs} & \textbf{Privacy Span Type} & \textbf{Privacy Spans Count} \\
            \cmidrule(lr){1-5}
             \rowcolor{lightgray!45} \multirow{4}{*}{Pri-DDXPlus} \cellcolor{white}
            & All   & 7759 & 149 & 46179 \\ 
            & Train & 5901 & 149 & 35583 \\
            & Dev   & 309  & 60  & 1659 \\
            & Test  & 1549 & 78  & 8937 \\ 
            \cmidrule(lr){2-5}  
            \rowcolor{lightgray!45} \multirow{4}{*}{Pri-NLICE}
            \cellcolor{white}& All   & 4062 & 64  & 18241 \\ 
            & Train & 3282 & 64  & 14933 \\
            & Dev   & 130  & 58  & 552 \\
            & Test  & 650  & 64  & 2756 \\ 
            \cmidrule(lr){2-5} 
            \rowcolor{lightgray!45} \multirow{4}{*}{Pri-SLJA}
            \cellcolor{white}& All   & 3901 & 142 & 10418 \\ 
            & Train & 3117 & 142 & 7980 \\
            & Dev   & 130  & 95  & 417 \\
            & Test  & 654  & 142 & 2021 \\ 
            \bottomrule  
        \end{tabular}
    }
    \vspace{-0.5em}
    \caption{The statistics of Pri-DDXPlus, Pri-NLICE and Pri-SLJA. Average privacy symptoms indicate the average privacy spans occur in one query.}
    \label{tbl:statistics}
\end{table*}

\section{Long-Tailed Distribution of Privacy Spans}
\label{app:long_dis}

In this section, we present the long-tail distribution of privacy spans, where most privacy spans are concentrated in the majority categories.
Here, a privacy span refers to \textbf{a specific description of a user's private information}, such as the description of symptoms, e.g., ``\textit{I’ve been having a persistent cough}''. 
The corresponding privacy span type indicates \textbf{the category of the private information}, such as the symptom type, e.g., ``\textit{cough}''.

Considering that we have three privacy-preserving datasets covering the medical and legal domains, we analyze the frequency of each privacy span type separately for each domain. \textbf{For the medical domain}, we plot the distribution of medical privacy spans in the Pri-DDXPlus and Pri-NLICE medical dataset, as shown in Figure \ref{fig:frequency_privacy_spans_1}. 
We observe that most medical privacy spans are concentrated on the top types, such as ``\textit{pain}'' and ``\textit{fever}''.
\textbf{For the legal domain}, we plot the distribution of legal privacy spans in Pri-SLJA legal dataset, as shown in Figure \ref{fig:frequency_privacy_spans_2}. 
We also observe that most legal privacy spans are concentrated on the top types, such as ``\textit{a person with full criminal responsibility}''.

Therefore, the privacy spans in both the medical and legal domains exhibit a long-tailed distribution, indicating that most privacy spans are concentrated in the majority types.

\begin{figure*}[!htbp]
  \centering
  \begin{minipage}{0.75\columnwidth}
    \centering
    \subfigure[Medical Privacy Spans Distribution]{
        \includegraphics[width=1\textwidth]
        {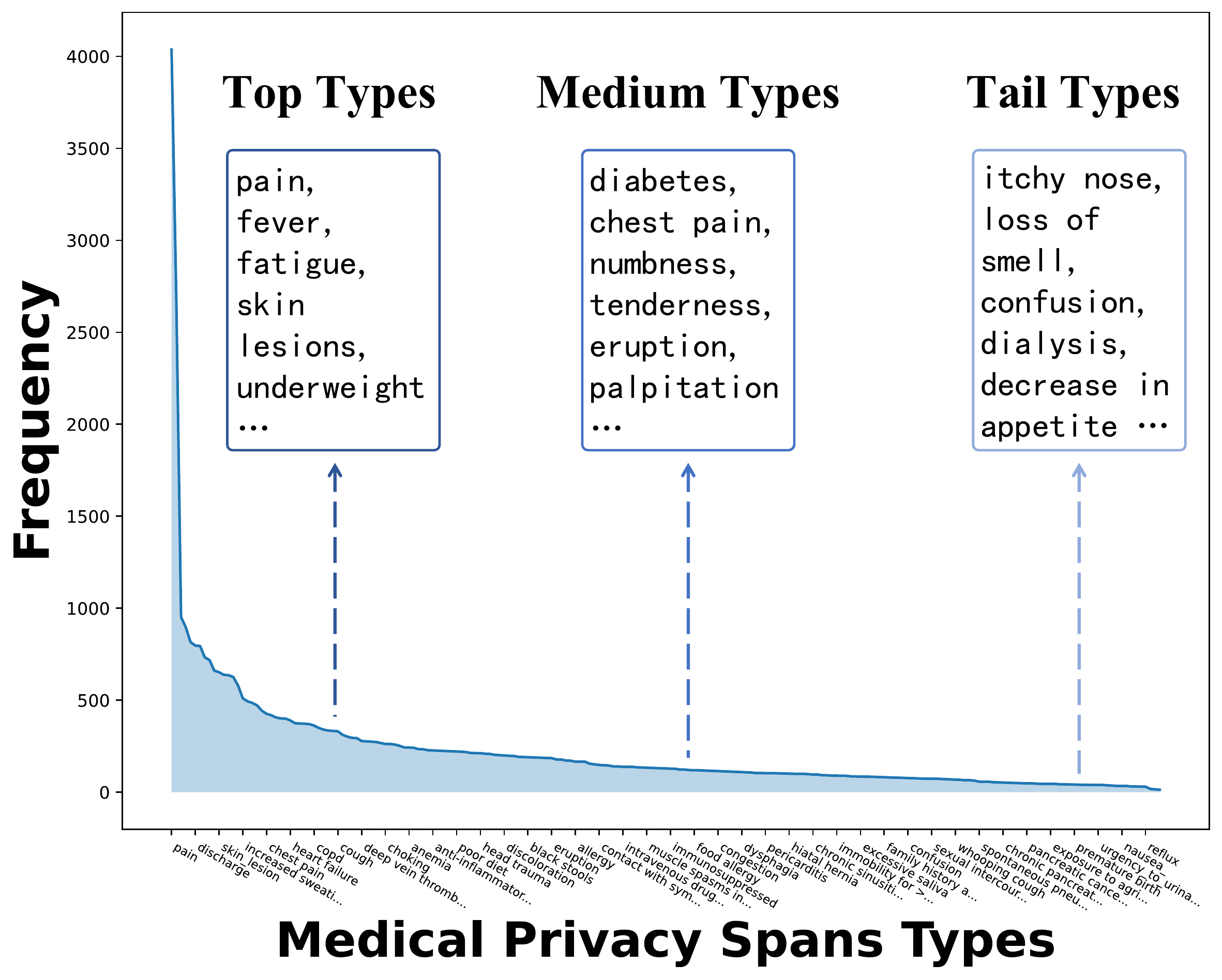}
        \label{fig:frequency_privacy_spans_1}
    }
  \end{minipage}%
  \begin{minipage}{0.75\columnwidth}
    \centering
    \subfigure[Legal Privacy Spans Distribution]{
        \includegraphics[width=1\textwidth]
        {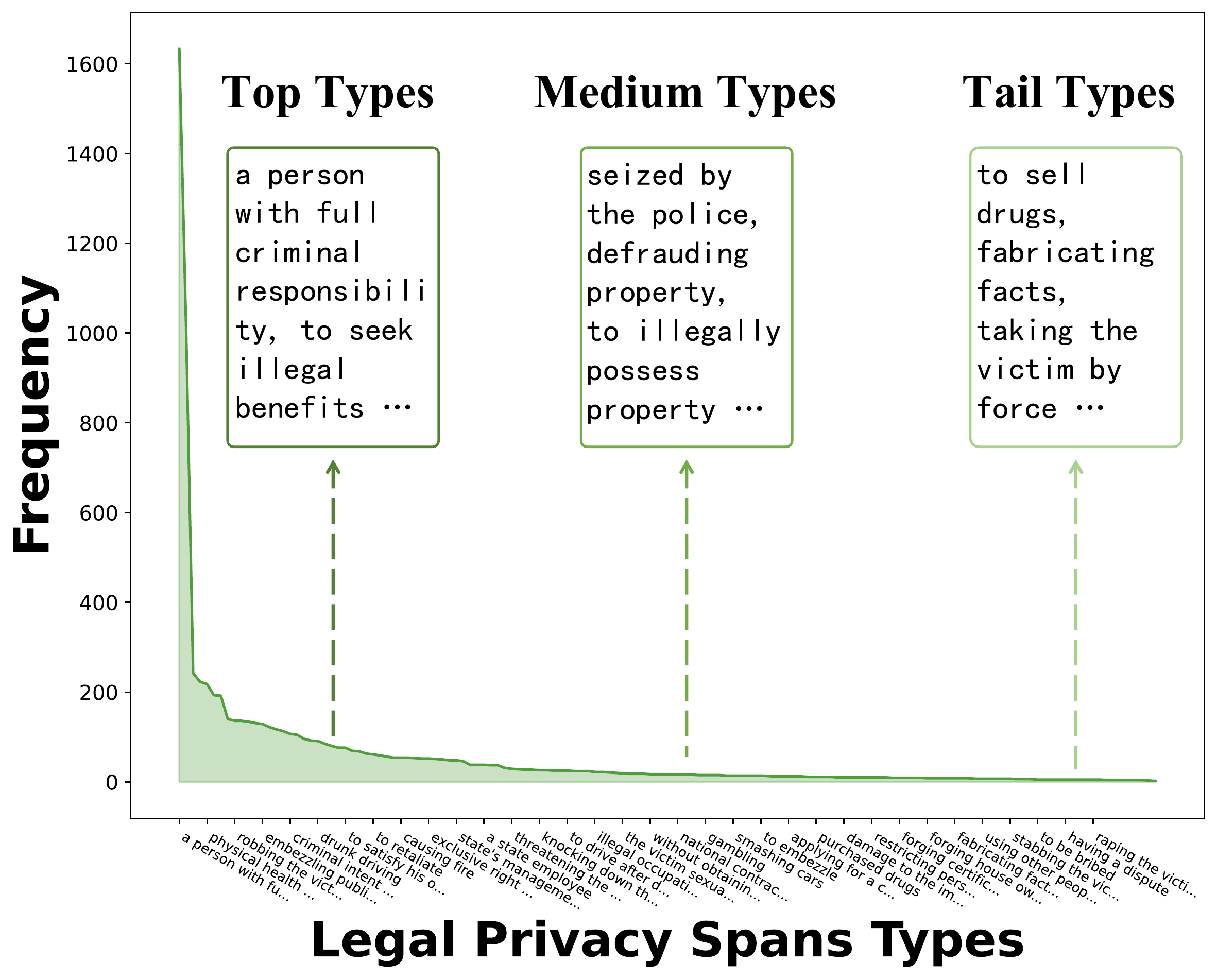}
        \label{fig:frequency_privacy_spans_2}
    }
  \end{minipage}
  \caption{Frequency distribution of privacy spans, highlighting the long-tail distribution where a small number of categories dominate the majority of occurrences.}
  \label{fig:frequency_privacy_spans}
\end{figure*}

\section{Experimental Setup Details}

\subsection{Evaluation Metrics}
\label{app:eva_metrics}
To fully evaluate the performance of different privacy-preserving methods, we focus on two aspects: inference performance and inference efficiency. 
We use \textbf{MC1, MC2, ROUGE-L, and LLM-J} to assess inference performance, and \textbf{Throughput (TP)} to evaluate inference efficiency.
The details of these metrics and their calculation methods are introduced as follows:

\paragraph{MC1/MC2:} 
We employ MC1 and MC2 \footnote{The code is available at \url{https://github.com/sylinrl/TruthfulQA}} \citep{zhang2024truthx} to measure the model's accuracy in selecting the correct answer among 4 options. 
We assign each sample in Pri-DDXPlus, Pri-NLICE, and Pri-SLJA with four options, including one correct answer and three incorrect ones. The details of the calculation process are as follows:

As for calculating MC1:
For each user input, we select the option with the highest probability as the model's choice.
MC1 is defined as the model's accuracy, which is calculated as the proportion of correctly answered inputs.

As for calculating MC2:
For each user input, we compute the normalized probability of the correct answer among the four options. 
The average of these normalized probabilities across all inputs is calculated as the MC2 score.

\paragraph{ROUGE-L:} 
We utilize ROUGE-L \citep{lin-2004-rouge} to assess the generation ability of different privacy-preserving methods.
ROUGE-L primarily measures the n-gram overlap between the reference text and the generated text. 
To evaluate the performance of these privacy-preserving methods, the reference text is the initial output without any privacy protection from the backbone LLM, while the generated text is the output with privacy protection.

\paragraph{LLM-Judge (LLM-J):}
As ROUGE-L primarily focuses on n-gram overlap between generated text and reference texts, which may not fully capture the semantic meaning or overall quality of the generated content, we further use the LLM-Judge (LLM-J)\citep{Lia2023judging} metric to assess the generation ability.
Specifically, we use the advanced LLM (i.e., GPT-4 \citep{openai2023gpt4}) to assess the quality of outputs considering relevance, clarity, and accuracy. 
The assessment prompt is shown in Appendix \ref{sec:eval_gene}. 
The LLM-J score ranges from 1 to 10, with higher scores indicating better quality. 

\paragraph{Throughput (TP):}
For inference efficiency, we use Throughput (TP), defined as the number of tokens generated per second, to evaluate the inference efficiency.
To ensure a fair comparison between different methods, we uniformly use sampling-based generation, as it effectively prevents privacy leakage from the generated outputs, as shown in Appendix \ref{app:proof_output_pro}. 
We set the sampling temperature to 1.0 and the maximum generation length to 1024.

\subsection{Compared Methods}
\label{app:com_method}
Here, we provide a more detailed introduction to all the compared methods, used to protect user privacy during LLM inference, including No Restoration, No Protection, $d_\chi$-privacy \citep{feyisetan2020privacy}, $d_\chi$-privacy on privacy spans, and Paraphrase \citep{justus2022limit, sai2023ldp}. 
The details are as follows:

\paragraph{No Restoration (lower bound):} This method involves transmitting user queries with privacy-sensitive spans removed, without attempting to restore the missing content on the server. 
As a result, this method serves as the performance lower bound among all privacy-preserving approaches. 
The degraded quality of the responses highlights the need for effective restoration techniques to bridge the gap between privacy protection and utility.

\paragraph{No Protection (upper bound):} In this method, user queries are transmitted directly to the server without any privacy protection or modifications. 
Since no information is removed or altered, the model operates on fully intact queries, achieving the best possible performance. 
Consequently, this method establishes the upper bound for all privacy-preserving techniques. 
The ROUGE-L score for No Protection is always 100.00, as the reference outputs for evaluation are from this method.

\paragraph{${d_\chi}$-privacy:}
As proposed by \citet{feyisetan2020privacy}, we can directly apply $d_\chi$-privacy mechanism to all tokens in the user query by injecting noise into the tokens' embeddings and replacing the initial tokens with their nearest counterparts. 
This prevents attackers from recovering the original tokens, thereby protecting privacy.

\paragraph{${d_\chi}$-privacy on privacy spans:}
Instead of applying $d_\chi$-privacy mechanism to the entire input, the client can only employ $d_\chi$-privacy only to the privacy spans in the user query, as the other parts of the query contain no privacy-sensitive information. 
This approach allows for a more appropriate and concise allocation of the privacy budget.

\paragraph{Paraphrase:}
According to \citet{justus2022limit, sai2023ldp}, the above methods, both applying $d_\chi$-privacy mechanism to tokens and achieving word-level privatization, suffer from the linear growth problem of the privacy budget.
They proposed to use generative models to paraphrase original inputs and achieve privacy protection similar to $d_\chi$-privacy.
Due to the client's computational resource limitations and to ensure a fair comparison with our method, we use the FLAN-T5-Base model \citep{hyu2024flant5} on the client side for paraphrasing in the Paraphrase baseline, as its model size is comparable to that of BERT-Base, which is used in our method.

\subsection{Implementation Details}
\label{app:imple_details}
We use Llama2-chat-7b \citep{touvron2023llama} as the LLM backbone on the server side, and BERT-base \citep{devlin2019bert} on the client side for weight estimation, as described in Section \ref{sec:inference}. 

During restoration vector training, the LLM parameters remain fixed, and we train the restoration vectors for 5 epochs with a batch size of 1.
The optimal number of edited heads $K$ is 175 for Pri-DDXPlus/Pri-SLJA and 125 for Pri-NLICE. 
The search process is shown in Section \ref{app:k}. 

During generation, we use a sampling-based decoding strategy with a temperature of 1.0 and a maximum generation length of 1024.
This is because sampling-based generation can effectively prevent privacy leakage from the generated outputs, as shown in Appendix \ref{app:proof_output_pro}. 
To evaluate the generation capabilities, we utilize GPT-4 \citep{openai2023gpt4} to assess the generated outputs.
The prompts are detailed in Appendix \ref{sec:gene_prompt}.

For the paraphrase baseline method, we employ the flan-t5-base model \citep{hyu2024flant5} on the client side, as its model size is comparable to BERT-base. 
Following \citet{justus2022limit}, we clip the final output logits between 0 and 1 during paraphrasing. 
As a result, the privacy budget for paraphrasing becomes $2n/\tau$, where $n$ represents the maximum length of the user query and $\tau$ is the generation temperature.

\section{Settings of Privacy Hyperparameter $\epsilon$}
\label{app:para}
According to \citet{feyisetan2019privacy} and the definition of $d_\chi$-privacy in Appendix \ref{app:method_preliminary_dx_privacy}, when applying the $d_\chi$-privacy mechanism to protect a single token, the privacy budget is $\epsilon d_e$ and $d_e$ is the maximum distance between any two token embeddings.
As proposed by \citet{justus2022limit}, with the length of input text increases, the privacy budget of $d_\chi$-privacy mechanism also grows linearly. 
Then, assuming that the the maximum length of the user query is $n$ and the maximum length of privacy spans in the user query is $n_{sp}$, \textbf{the privacy budget of $d_\chi$-privacy is $n\epsilon$} and \textbf{the privacy budget of $d_\chi$-privacy on privacy spans is $n_{sp}\epsilon$}.
In addition, as pointed by \citet{justus2022limit, sai2023ldp}, \textbf{the privacy budget of paraphrase method is $2n/\tau$}, where $\tau$ is the generation temperature used during paraphrasing, and $n$ represents the maximum length of user queries. 

\textbf{The privacy budget of PrivacyRestore is $2\epsilon$}, according to Theorem \ref{theo_2}.
To ensure the same privacy budget for a fair comparison, we need to determine the values of different hyperparameters for different methods on different datasets, such as  $\epsilon$ for $d_\chi$-privacy (on privacy spans), PrivacyRestore and $\tau$ for paraphrase.

Firstly, We set the privacy hyperparameter $\epsilon$ to 75.00 for PrivacyRestore.
Next, we compute the maximum of the users' inputs lengths $n$, privacy spans lengths $n_{ps}$, and distances between word embeddings $d_e$ across three privacy-preserving datasets.
Then, we calculate the corresponding $\epsilon$ for $d_\chi$-privacy (on privacy spans) and $\tau$ for paraphrase to \textbf{control the overall privacy budget at 150}, as detailed in Table \ref{tbl:para}.

\begin{table*}[!ht]
    \centering
    \vspace{-6pt}
    \def\arraystretch{0.95}
    \resizebox{0.97\textwidth}{!}{
        \begin{tabular}[h]{c 
        ccc  ccc  cc  c  c
        }
            \toprule
            
            \multirow{2}{*}{\textbf{Datasets} }&
            \multicolumn{3}{c }{\textbf{$d_\chi$-privacy}} &
            \multicolumn{3}{c }{\textbf{$d_\chi$-privacy on privacy spans}} &
            \multicolumn{2}{c }{\textbf{Paraphrase}} & 
            \multicolumn{1}{c }{\textbf{PrivacyRestore}} & \multirow{2}{40pt}{ \textbf{Privacy Budget} } \\
            
            \cmidrule(lr){2-4} \cmidrule(lr){5-7} \cmidrule(lr){8-9} \cmidrule(lr){10-10}
            ~ & 
           \multicolumn{1}{p{30pt}}{\centering $n$} & 
           \multicolumn{1}{p{30pt}}{\centering $d_e$} & 
            \multicolumn{1}{p{30pt}}{\centering \textbf{$\epsilon$}} &
            
            \multicolumn{1}{p{40pt}}{\centering $n_{sp}$} & 
            \multicolumn{1}{p{40pt}}{\centering $d_e$} & 
            \multicolumn{1}{p{40pt}}{\centering \textbf{$\epsilon$}} & 
            $n$ & $\tau$ &
             \multicolumn{1}{p{80pt}}{\centering $\epsilon$} \\
            \cmidrule(lr){1-11}
            \addlinespace[5pt] 
          Pri-DDXPlus & 106.00 & 1.64 & \textbf{0.86} &  49.00 & 1.64 & \textbf{1.86}  & 106.00 & \textbf{1.41}  & \textbf{75.00} & \underline{\textbf{150}} \\
            \addlinespace[5pt] 
          Pri-NLICE & 72.00  & 1.39  & \textbf{1.50} & 38.00  & 1.39  & \textbf{2.84}   & 72.00  & \textbf{2.08}   & \textbf{75.00} & \underline{\textbf{150}} \\
            \addlinespace[5pt] 
          Pri-SLJA & 193.00 & 1.45 & \textbf{0.54} &  42.00 & 1.45 & \textbf{2.46}  & 193.00 & \textbf{0.78} & \textbf{75.00} & \underline{\textbf{150}} \\
          \addlinespace[2pt] 
          \bottomrule \\
          
        \end{tabular}
    }
    \vspace{-0.7em}

     \caption{The settings of privacy hyperparameters for different baselines across all privacy-preserving datasets.}
    \label{tbl:para}
\end{table*}

  


\section{Additional Baselines}
\label{app:new_baselines}
In addressing the challenge of safeguarding user privacy during LLM inference, recent studies have explored innovative approaches that leverage small language models to either anonymize or substitute private information within user queries. 
To evaluate the efficacy of our approach, we implemented two baselines from recent studies on the Pri-DDXPlus dataset. 
(a). \textbf{LLM-anonymization}, proposed by \citet{StaabLLMAnonymization}, which uses a language model to anonymize text by repeatedly removing personal attributes identified by an adversarial inference model. (b). \textbf{IncogniText}, introduced by \citet{FrikhaIncogniText}, which anonymizes text by iteratively using an adversarial model to identify private attribute inferences and an anonymization model to rewrite the text, misleading potential attackers into predicting incorrect private attribute values while preserving text utility.


As shown in Table \ref{tbl:new_baselines}, our method significantly outperforms LLM-anonymization and IncogniText on Pri-DDXPlus dataset, strongly validating the effectiveness of our approach.


\begin{table}
	\centering
        \def\arraystretch{1.1}
	\resizebox{1\columnwidth}{!}{
       \begin{tabular}{ l |  ccccc  }
	   \toprule
            \textbf{Methods}  & \textbf{MC1} $\uparrow$ & \textbf{MC2} $\uparrow$ & \textbf{ROUGE-L} $\uparrow$ & \textbf{LLM-J} $\uparrow$  \\
            \cmidrule(lr){1-5} 
             LLM-anonymization &
            52.09($\downarrow${10.88}) & 49.94($\downarrow${10.25}) &25.22($\downarrow${2.02}) & 2.61($\downarrow${1.86})  \\
             IncogniText &
            55.26($\downarrow${7.71}) & 53.73($\downarrow${6.46}) & 25.94($\downarrow${1.30}) & 3.85($\downarrow${0.62})  \\
            \rowcolor{lightgray!45}
             \textbf{PrivacyRestore} & \textbf{62.97} & \textbf{60.19} & \textbf{27.24} & \textbf{4.47}\\
            \midrule
		\end{tabular}
	}
\caption{
Comparison of the performance and the inference efficiency between PrivacyRestore, \textbf{LLM-anonymization} and \textbf{IncogniText} methods across Pri-DDXPlus dataset. 
The downward arrow in the table indicates the performance gap of two baseline methods compared with PrivacyRestore.}
\label{tbl:new_baselines}
\end{table}

\section{Details of Privacy Protection Evaluation}
\label{app:detail_PP_eval}
In this section, we provide more details on our implementation of embedding inverse attack \citep{hao2023sentence, john2023text} and attribute inference attack \citep{hao2022you} to evaluate the privacy protection performance of different privacy-preserving baselines and our method.
Lower attack performance indicates stronger privacy protection provided by these methods.

\paragraph{Embedding Inverse Attack:} 
As proposed by \citet{hao2023sentence, john2023text}, embedding inversion attacks aim to recover user privacy from the embeddings of user inputs. 
Specifically, a generative model (e.g., GPT-2 model \citep{radford2019language}) is used to generate the user's private information based on the given embedding.
We implement embedding inversion attacks for the privacy-preserving baselines and our method to evaluate their privacy protection performance.
The implementation details are as follows:

We use the gpt2-medium model \citep{radford2019language} as the generative model, employing greedy search during generation and setting the maximum generation length to 256.
\textbf{For PrivacyRestore}, the client transmits the incomplete user query and the meta vector. 
The incomplete user query does not contain any user privacy after removing the privacy spans and is secure for the user.
The meta vector contains the information of privacy spans and we perform embedding inverse attack on the meta vector. 
We use a fully-connected layer to transform the meta vector's dimension to the dimension of hidden state of GPT-2 model.
Then we directly input the transformed meta vector as the input embedding.
We fine-tune the GPT-2 model and the fully connected layer simultaneously, on the training set for 20 epochs, using a learning rate of 1e-5.
\textbf{For $d_\chi$-privacy (on privacy spans) and paraphrase}, the client only transmits the garbled user query after applying the $d_\chi$-privacy mechanism or paraphrasing. 
We then perform the embedding inverse attack on the garbled user query to recover the privacy spans.
Here, we do not need to transform the dimension and can directly input the garbled user query as the input context for the GPT-2 attack model.
Then attack model can recover the privacy spans according to the garble user query.
We finetune the GPT-2 model on the training set for 20 epochs using the learning rate of 1e-5.

To evaluate the attack's performance, we compute the ROUGE-L score between the generated output of the attack model and ground true privacy spans in the user query, where higher scores indicate better attack effectiveness.

\paragraph{Attribute Inference Attack:} 
According to \citet{hao2022you}, attribute inference attack attempts to infer user's private attribute even when the user query is protected by some privacy-preserving methods.
In our scenario, we use attribute inference attacks to infer the privacy spans in the user query.
The implementation details are as follows:

Following \citet{hao2022you}, we construct a multi-layer perceptron (MLP) as the classifier, with the output dimension corresponding to the entire vocabulary size. 
We use the classifier to predict the token IDs of the privacy spans in the user query. 
Since the query contains multiple privacy spans, and each span consists of multiple tokens, this classification task is a multi-label classification.
\textbf{For PrivacyRestore}, we also perform attribute inference attacks on the meta vector, so the input dimension of the classifier corresponds to the dimension of the meta vector.
We finetune the classifier on the training set for 20 epochs using the learning rate of 1e-5.
\textbf{For $d_\chi$-privacy (on privacy spans) and paraphrase}, we perform attribute inference attack on the garbled user query.
We utilize GPT-2 model \citep{radford2019language} to process the query and obtain the last token's hidden state as the vector representation. 
Classification is then performed on this hidden state.
We finetune the classifier and the GPT-2 model jointly, on the training set for 20 epochs using the learning rate of 1e-5.

To evaluate the attack's performance, we calculate the F1 score of the classification, where a higher F1 score indicates a more successful attack.

\section{More Privacy Protection Evaluation Results}
\label{app:more_protection_evaluation_results}





\subsection{Concatenated Text Attack}

In Section \ref{sec:pri_protect_eval}, the implementation of embedding inverse attack follows previous work \citep{hao2023sentence}, which merely takes meta vectors derived from privacy spans as input. 
This approach, however, may overlook the contextual information in the incomplete user query.
Therefore, we propose the Concatenated Text Attack by firstly using embedding inverse attack to transform the meta vector to the text format and then concatenate it with the incomplete user query to add more contextual information for recovering privacy spans. 
The implementation details are as follows: 

We finetune two attack models: one to transform the meta vector into text format, and the other to recover privacy spans from the concatenated text.
For the first model, we finetune a GPT-2 model, where the input is the meta vector and the output is the privacy spans, similar to the embedding inverse attack process.
Then we concatenate the generated output from the first model with the incomplete user query as the input to the second model.
As for the second model, we also finetune a GPT-2 model which aims to utilized the incomplete user query to improve the quality of the generated output from the first attack model.
For both attack models, we finetune them on the train set for 20 epochs using the learning rate of 1e-5.
We also utilized the ROUGE-L scores between the recovered results and the privacy spans as the evaluation metric.

The experimental evaluation of the Concatenated Text Attack is presented in Table \ref{tbl:concatenated_text_attack}. 
The experiment results show that, although unifying the vector format of the meta vector and the text format of incomplete user query, the attack performance for our method is still poor, demonstrating the effectiveness of our method. 

\begin{table}[h]
  \centering
  \def\arraystretch{1.2}
  \resizebox{0.48\textwidth}{!}{
  \begin{tabular}{l  c c c c c c }
    \toprule
    \textbf{$\epsilon$ values} & \textbf{1} & \textbf{20} & \textbf{40} & \textbf{75} & \textbf{125} & \textbf{175} \\
    \cmidrule(lr){1-7}  
    Pri-DDXplus   & 0.0112  &0.0107  &0.0130  &0.0093 & 0.0115  &0.0024 \\
    \cmidrule(lr){2-7}
    Pri-NLICE & 0.0566 & 0.0486 & 0.0427  &0.0467 & 0.0423 & 0.0350 \\
    \cmidrule(lr){2-7}
    Pri-SLJA   &0.0027  &0.0011 & 0.0022 & 0.0024 & 0.0021 & 0.0028 \\
    \toprule
  \end{tabular}
  }
  \caption{ROUGE-L Scores for Concatenated Text Attack Across Different $\epsilon$ Values}
   \label{tbl:concatenated_text_attack}
\end{table}

\subsection{Simulating Activation Steering Attack}
We assume the attacker is aware that the meta vector will be used for activation steering on the server for information restoration.
\textbf{The attacker can also simulate activation steering while recovering the privacy spans in the user query}.
Considering the LLM weights on the server are kept secret, the attack only can conduct the activation steering on the other generative model, such as GPT-2 \citep{radford2019language} model. 
The implementation details are as follows:

First, due to the heterogeneity between the attack model (GPT-2) and the LLM on the server (Llama-2-7b), we use a fully connected layer to transform the meta vector’s dimension to fit in the attack model. 
Specifically, since the meta vector is applied to the head output and its initial dimension matches the head output of the LLM, the fully connected layer adjusts it to the head output dimension of the attack model (GPT-2).
Next, we input the incomplete user query into the attack model and use the adjusted meta vector to perform activation steering, prompting the model to generate the privacy spans in the query.
We fine-tune the GPT-2 model and the fully connected layer jointly for 20 epochs with a learning rate of 1e-4.
We utilized the ROUGE-L scores between the recovered results and the privacy spans as the evaluation metric.

As shown in Table \ref{tbl:simulate_activation_steering_attack}, the Simulating Activation Steering Attack demonstrated limited performance across various $\epsilon$ values on all three datasets. 
This weakness may be attributed to that the meta vector are trained offline for the server’s LLMs.
Although we have used fully connected layer to transform the dimension of the meta vector, applying the meta vector to the attack model still leads to incompatibility.
\begin{table}[H]
  \centering
  \def\arraystretch{1.2}
  \resizebox{0.48\textwidth}{!}{
  \begin{tabular}{l  c c c c c c }
    \toprule
    \textbf{$\epsilon$ values} & \textbf{1} & \textbf{20} & \textbf{40} & \textbf{75} & \textbf{125} & \textbf{175} \\
    \cmidrule(lr){1-7}  
    Pri-DDXplus   &  0.0023&0.0329&0.0321&0.0329&0.0310&0.0365 \\
    \cmidrule(lr){2-7}
    Pri-NLICE & 0.0165&0.0123&0.0118&0.0170&0.0283&0.0315 \\
    \cmidrule(lr){2-7}
    Pri-SLJA  & 0.0161&0.0818&0.0862&0.0861&0.1048&0.1059 \\
    \toprule
  \end{tabular}
  }
  \caption{ROUGE-L Scores for Simulating Activation Steering Attack Across Different $\epsilon$ Values}
   \label{tbl:simulate_activation_steering_attack}
\end{table}

\subsection{Hidden State Attack}
Hidden State Attack\citep{carlini2021extract} attempts to perform a training data extraction attack to recover individual training examples by querying the language model. We have implemented a Hidden State Attack employing the LLaMA-7B architecture as the designated attack model. The objective of this attack was to infer private information from steered hidden states, specifically targeting the first, sixteenth, and final layers of the network. The efficacy of the attack was quantitatively assessed using the ROUGE-L metric, which measures the lexical similarity between the output generated by the attack model and the original user query containing sensitive private information.

As indicated by the empirical results presented in Table\ref{tbl:hidden_state_attack}, the attack directed at the final layer demonstrated a marginal improvement in performance when compared to attacks executed on the first and sixteenth layers. This observation can be attributed to the fact that hidden states within the final layer are fully restored, in contrast to the earlier layers where only partial restoration is achieved. Nevertheless, it is crucial to emphasize that the performance of all conducted attacks remained notably weak. This outcome underscores the robustness of our proposed methodology in safeguarding user privacy. Furthermore, it is pertinent to state that our method does not introduce privacy vulnerabilities at the model level, as noise has been incorporated into the meta vector, thereby obfuscating sensitive information.
\begin{table}[H]
  \centering
  \def\arraystretch{1.1}
  \resizebox{0.48\textwidth}{!}{
  \begin{tabular}{l  c c c }
    \toprule
    \textbf{$\epsilon$ values} & \textbf{0.01} & \textbf{0.86} & \textbf{2.01}  \\
    \cmidrule(lr){1-4}  
    Hidden State Attack on 1st layer   &  0.0023&0.0329&0.0321 \\
    \cmidrule(lr){2-4}
    Hidden State Attack on 16th layer & 0.0165&0.0123&0.0118\\
    \cmidrule(lr){2-4}
    Hidden State Attack on final layer  & 0.0161&0.0818&0.0862 \\
    \toprule
  \end{tabular}
  }
  \caption{ROUGE-L Scores for Hidden State Attack on Different Layers Across Different $\epsilon$}
   \label{tbl:hidden_state_attack}
\end{table}

\section{Analysis of Output Privacy Protection}
\label{sec:output_privacy_spans}
In this section, we evaluate the privacy leakage in the generated output of our method by implementing Embedding Inversion Attacks (EIA) and Attribute Inference Attacks (AIA). 
We also directly count the frequency of privacy span occurrences in the generated outputs.
The details of these attack methods are as follows:

\paragraph{Embedding Inverse Attack for the generated output.}
Embedding inversion attacks \citep{hao2023sentence, john2023text} directly utilize the generative model (e.g., GPT-2) to generate privacy spans in the user query based on the attacked embedding.
Although the generated output is in text format rather than embedding format, we still input it into the GPT-2 model to generate the privacy spans from the user query.

To be specific, we utilize the GPT-2 model \citep{radford2019language} as the generative model and set the maximum generation length to 256.
The input of the GPT-2 attack model is the generated output and the target output is the privacy spans in the user query.
We finetune the GPT-2 model on the training set for 20 epochs using the learning rate of 1e-5.
To evaluate attack performance, we compute the ROUGE-L score between the output generated by the attack model and the ground truth privacy spans in the user query.

\paragraph{Attribute Inference Attack for the generated output.}
Attribute inference attack \citep{hao2022you} attempts to steal user privacy by performing classification on the generated output, where the target labels corresponding to the token IDs of those privacy spans.
Since each user query contains multiple privacy spans and each privacy span contains multiple tokens, this classification task is naturally a multi-label classification task. 

First, we use the GPT-2 model \citep{radford2019language} to process the text input and obtain the hidden state of the last token as its vector representation. 
Next, following \citet{hao2022you}, we construct a multi-layer perceptron (MLP) model as the classifier. The classifier's input is the vector representation, and the output dimension corresponds to the vocabulary size.
We finetune the GPT-2 model along with the MLP on the training set for 20 epochs, using a learning rate of 1e-5.
To evaluate the attack performance, we compute the F1 score of the classification results, where a higher F1 score indicates a more successful attack.





\section{Details of Privacy Protection Robustness for Long Queries}
\label{app:long_input}
In this section, we will provide more implementation details and experiment results analysis when evaluating the privacy protection robustness of $d_\chi$-privacy and our method.

\subsection{Different Protected Text Length for $d_\chi$-privacy}
\label{app:long_input_1}
As shown in Section \ref{sec:longer_q}, we randomly select a proportion of token in user query to simulate the protected text and larger proportion indicates longer protect text.
The proportion of selected token is denoted as the \textbf{$d_\chi$-privacy Percentage}. 
As presented by \citet{feyisetan2019privacy, feyisetan2020privacy}, the $d_\chi$-privacy mechanism protects input by injecting noise into the token embeddings and replacing the original tokens with their nearest neighbors. 
To attack the garbled query, we implement two types of attacks: prompt injection attack \citep{ xuchen2024sign} and attribute inference attack \citep{hao2022you}, both commonly used for attacking text inputs.
The details of implementation of these two attack methods are as follows:

\textbf{For prompt injection attack}, following \citet{xuchen2024sign}, we add extra instructions before and after the garbled query, to prompt the LLM in the server to output the protected text instead of following the initial user query.
And then we intercept the output returned by the LLM on the server for user privacy.
The template for the additional instructions is provided in Appendix \ref{app:prompt_PI}.
To evaluate the attack performance, we calculate the ROUGE-L score between the returned output and the protected text.
A higher ROUGE-L score indicates greater overlap between the returned output and the protected text, signifying more successful attack results.
\textbf{For attribute inference attack}, inspiring by \citet{hao2022you}, We firstly utilize GPT-2 model \citep{radford2019language} to process the garbled query and obtain the last token's hidden state. 
Next, we construct a multi-layer perceptron (MLP) as the classifier to classify the hidden states, with the target labels being the token IDs of the protected text. 
This is a multi-label classification task.
We finetune the classifier and the GPT-2 model on the training set for 20 epochs using the learning rate of 1e-5.
The attack performance is evaluated using the classification F1 score.

As shown in Figure \ref{fig:pi_dx} and Figure \ref{fig:at_dx}, the attack performance of prompt injection attack and attribute inference attack across all three datasets are all grows with the larger $d_\chi$-privacy percentage.
These experiment results reflect the linear growth problem of privacy budget in $d_\chi$-privacy.

\subsection{Different Protected Text Length for PrivacyRestore}
\label{app:long_input_2}
For PrivacyRestore, we randomly choose a proportion of privacy spans in the user query as the protected text and the proportion is denoted as the \textbf{Privacy Span Ratio $\alpha$}.
Larger $\alpha$ indicate the longer protected text.
Considering that, in our method, the client only transmits the incomplete query with the meta vector and the incomplete query contains no privacy information, then we implement embedding inverse attack \citep{hao2023sentence, john2023text} and attribute inference attack \citep{hao2022you} on the meta vector across different $\alpha$ values. 
The details of implementation of these two attack methods are as follows:

\textbf{For embedding inverse attack}, we firstly fully-connected layer to transform the meta vector's dimension to the dimension of hidden state of GPT-2 attack model.
Then we directly input the transformed meta vector as the input embedding to the attack model, prompting it to generate the privacy spans in the user query.
We finetune the fully-connected layer with the GPT-2 attack model on the training set for 20 epochs using the learning rate of 1e-5.
The attack performance is assess by the ROUGE-L score between the generated output from the attack model and the protected text.
\textbf{For attribute inference attack}, we construct a multi-layer perceptron (MLP) as the classifier to classify the meta vector, with the target labels being the token IDs of the protected text. 
This is also a multi-label classification task.
We finetune the classifier on the training set for 20 epochs using the learning rate of 1e-5.
The attack performance is evaluated using the classification F1 score.

As shown in Figure \ref{fig:ea_pr} and \ref{fig:ai_pr}, the ROUGE-L score for the embedding inverse attack remains nearly stable across different $\alpha$ values in the Pri-SLJA and Pri-DDXPlus datasets. 
What's a little strange is the ROUGE-L score in the Pri-NLICE dataset shows a slight increase.
The possible reason is that higher ratio indicating more privacy spans and \textbf{resulting longer reference string when compute the ROUGE-L score}.
Since ROUGE-L measures the overlap between the generated output and the reference string, a longer reference string may slightly boost the score.
The F1 score for the attribute inference attack remains stable across all three datasets. 
The stable performance in both attack scenarios provides empirical support for Theorem \ref{theo_2}.
Our method effectively and inherently solves the linear growth problem of the privacy budget, achieving robust and stable privacy protection performance regardless of the length of the protected text, even with long protected text.

\section{Details of Evaluation of Handling Out-of-Set Privacy Spans}
\label{sec:new_privacy_spans}
In this section, we will evaluate out method when handling those out-of-set privacy spans.
As shown in Figure \ref{fig:frequency_privacy_spans}, most of privacy spans focus on the majority categories.
Our core set of predefined privacy spans easily covers the majority of categories, even though it cannot cover all privacy span types.
To evaluate the performance of our method when the core set cannot cover all privacy span types, we assume that the core set contains only the top 5, 40, 80, 100, or 120 privacy span types and assess our method. 
Additionally, we provide results when the core set covers all 149 privacy span types in the Pri-DDXPlus dataset.

As shown in Table \ref{table:top_predefine}, our approach outperforms the No Restoration baseline, with performance gains increasing as the predefined span set expands. 
Notably, \textbf{even when limited to the top 100 types, our method achieves significant improvements across multiple metrics}. 
These findings highlight the robustness and efficiency of our method in handling those out-of-set privacy spans when our predefined cores set cannot cover all privacy spans.

\begin{table*}[t]
\centering
\def\arraystretch{1.1} 
\resizebox{0.82\textwidth}{!}{
\large
\begin{tabular}{l cccc}
\hline
 \textbf{Methods$\uparrow$} & \textbf{MC1$\uparrow$} & \textbf{MC2$\uparrow$} & \textbf{RL$\uparrow$} & \textbf{LLM-J$\uparrow$}\\ 
\hline
 No Restoration (lower bound) & 33.57 &  32.49& 25.19& 3.21 \\ 
Predefine only top 5 & 38.21 &  36.17& 25.82& 3.57 \\ 
Predefine only top 40 & 44.28 &  42.00& 25.89& 3.83 \\ 
Predefine only top 80 & 45.83 &  43.53& 26.59& 3.95 \\ 
\rowcolor{lightgray!45}
\textbf{Predefine only top 100} & \textbf{54.93${(\uparrow21.36)}$} &  \textbf{52.15${(\uparrow19.66)}$}& \textbf{26.37${(\uparrow1.18)}$}& \textbf{4.19${(\uparrow0.98)}$} \\ 
Predefine only top 120 & 58.42 &  55.40& 26.87& 4.27 \\ 
Predefine only all (top 149) & 62.97${(\uparrow29.40)}$ &  60.19${(\uparrow27.70)}$ & 27.24${(\uparrow2.05)}$& 4.47${(\uparrow1.26)}$ \\ 
\bottomrule
\end{tabular}
}   
\caption{Performance comparison across different predefined privacy span type sets $\mathcal{C}$ in Pri-DDXPlus}
\label{table:top_predefine}
\end{table*}

\section{Details of Extension for Users Unable to Determine Privacy Spans}
\label{app:un_deter_privacy}
In this section, we evaluate the performance of combining PrivacyRestore with existing text sanitization techniques \citep{Kan2023ProtectingUP,Chen2023HideAS} to address the situation where users cannot or are unwilling to determine privacy spans themselves.
In our main setting, we follow the principle of ``Information Self-Determination Right'' and assume that the user should determine the privacy spans in their queries by themselves.
However, we also consider the situation when the user cannot or is unwilling to identify the privacy spans.
Thanks to our method is totally orthogonal to the existing Text Sanitization techniques \citep{Kan2023ProtectingUP,Chen2023HideAS}, we can use text sanitization technique to identify and remove privacy spans automatically and restore information during LLM inference by our method.

Specifically, the pipeline of combining text sanitization technique and our method consists of three stages: \textbf{Privacy Spans Identification}, \textbf{User Query Sanitization} and \textbf{PrivacyRestore}. The details of these three stages are as follows:

\paragraph{Privacy Spans Identification:} 
Following \citet{Kan2023ProtectingUP,Chen2023HideAS}, we construct a classifier based on the BERT-base-uncased model \cite{devlin2019bert}. 
The input to the classifier is the user query, and the target labels are the types of privacy spans in the query. 
Considering that each query contain multiple privacy spans and this is a multi-label classification task.
We use the classifier to identify the types of privacy spans present in the user query.
We finetune the classifier on the training set for 10 epochs using the learning rate of 1e-4.
To evaluate the identification performance, we compute the precision, recall and F1 score of the classification.

As shown in Table \ref{table:spans_identification}, the classification results of our classifier are superior, achieving an F1 score of 99.66.

\begin{table}[H]
\centering
\def\arraystretch{1.5} 
\resizebox{0.48\textwidth}{!}{
\large
\begin{tabular}{l ccc}
\hline
 & \textbf{Precision} & \textbf{Recall} & \textbf{F1} \\ 
\hline
Privacy Spans Identification & 99.16$_{\pm 0.21}$ & 98.78$_{\pm 0.31}$ & 99.66$_{\pm 0.27}$ \\ 
\bottomrule
\end{tabular}
}   
\caption{Privacy Spans Identification accuracy. The results of the three experiments are presented, with the variance displayed in subscript.}
\label{table:spans_identification}
\end{table}

\paragraph{User Query Sanitization:} 
After identifying all privacy spans in the user query, we need to remove all these privacy spans from the user query to achieve sanitization. 
Inspiring by \citet{Kan2023ProtectingUP,Chen2023HideAS}, we finetune a Qwen-2.5-0.5B model \cite{qwen2} to conduct the text sanitization.
Specifically, the model takes the user query and the identified privacy span types as input and outputs a sanitized version of the user query with the privacy spans removed.
We finetune the Qwen-2.5-0.5B model on the train set for 15 epochs using the learning rate of 1e-5. 

To evaluate the efficacy of the text sanitization, we conducted both Attribute Inference Attacks(AIA) and Embedding Inversion Attacks(EIA) on the sanitized queries.
As shown in Table \ref{table:sanitized_attack}, the performance of both attack methods are very low, demonstrating that our sanitization method can effectively protect the user privacy.

\begin{table}[H]
\centering
\def\arraystretch{1} 
\resizebox{0.45\textwidth}{!}{
\begin{tabular}{l cc}
\hline
 & \textbf{EIA (ROUGE-L)} & \textbf{AIA (F1)} \\ 
\hline
No Protection & $0.40$ & $0.70$ \\ 
Sanitized Results& $0.06{(\downarrow0.34)}$&$0.07{(\downarrow0.63)}$\\
\bottomrule
\end{tabular}
}   
\caption{Attack results on sanitized queries. EIA refers to the embedding inverse attack, with the evaluation metric being ROUGE-L. AIA denotes the attribute inference attack, evaluated using the F1 score.}
\label{table:sanitized_attack}
\end{table}


\paragraph{PrivacyRestore:} 
Following the text sanitization, we use PrivacyRestore to restore the information during LLM inference on the server.
We present the performance results of our method when user can determine privacy spans (\textbf{PR+PR}), combining our method with text sanitization (\textbf{PR+TS}), only using text sanitization (\textbf{TS only}) and the No Restoration baseline in  Table \ref{table:prires_sanitized}.

As the experiment results show, even in scenarios where users are unable to identify privacy spans, the combination of our method with text sanitization (PR+TS) results in a significant enhancement in performance compared to the No restoration baseline (lower bound) and only using text sanitization (only TS).
The utility performance achieved is notably superior, suggesting that our method is effective in preserving privacy while simultaneously optimizing utility.
Moreover, the performance metrics of combining our method with text sanitization are comparable to those when the user can determine privacy spans themselves (PR+PS).
This comparison further underscores the robustness of combining our method with text sanitization and validates the efficacy of our approach in real-world applications, even when users cannot determine privacy spans themselves.


\begin{table}[pt]
\centering
\def\arraystretch{1.5} 
\resizebox{0.48\textwidth}{!}{
\begin{tabular}{l cccc}
\hline
 & \textbf{MC1$\uparrow$} & \textbf{MC2$\uparrow$} & \textbf{RL$\uparrow$} & \textbf{LLM-J$\uparrow$} \\ 
\hline
No Restoration & 33.57  & 32.49  & 25.19 & 3.21 \\
TS only & 29.63$(\downarrow 3.94)$ & 30.85$(\downarrow 1.64)$ & 25.45$(\uparrow 0.26)$ & 3.46$(\uparrow 0.25)$ \\
PR+PS & 62.97$(\uparrow 29.40)$ & 60.19$(\uparrow 27.70)$ & 27.24$(\uparrow 2.05)$ & 4.47$(\uparrow 1.26)$ \\
\rowcolor{lightgray!45}
\textbf{PR+TS} & \textbf{62.87}$(\uparrow 29.30)$ & \textbf{59.97}$(\uparrow 27.48)$ & \textbf{26.47}$(\uparrow 1.28)$ & \textbf{4.28}$(\uparrow 1.07)$ \\ 
\bottomrule
\end{tabular}
}
\caption{The performance of combining our method with text sanitization technique. \textbf{TS only} indicates only use sanitization methods without combining PrivacyRestore. \textbf{PR+PS} indicates PrivacyRestore when the user can determine privacy spans by themselves. 
\textbf{PR+TS} denotes combining PrivacyRestore and text sanitization to address the situation when the user cannot identify privacy spans by themselves. 
Three methods are compared with \textbf{No Restoration} baseline (lower bound).}
\label{table:prires_sanitized}
\end{table}

\section{Details to Ablation Study}
\label{app:abla}
In this section, we conduct additional experiments to analyze the impact of the number of edited heads and evaluate the performance of our method across varying LLM backbones.

\subsection{Hyperparameter Analysis of the Number of Edited Heads}
\label{app:k}
We evaluate the performance of our methods using different numbers of edited heads, $K$, across the development sets of three privacy-preserving datasets.
For simplicity, we compute MC2 to represent classification performance, LLM-J to measure generation performance, and TP to indicate inference efficiency.

As shown in Table \ref{tbl:vary_heads}, according to the MC2 score, the optimal value of $K$ is 175 for the Pri-DDXPlus and Pri-SLJA datasets, and 125 for the Pri-NLICE dataset.
The performance degradation as $K$ increases can be attributed to the cumulative effect of multiple edited heads. As more heads are modified, the activations progressively deviate from their initial values, potentially compromising the LLM's general capabilities.
Moreover, throughput increases with larger $K$ because we need to inject the meta vector for each head in $\mathcal{H}_k$ using Eq \ref{eq:restore} on the server. Consequently, more heads indicate more injections, which increases the inference time on the server.

\begin{table*}[!htbp]
  \centering
  \resizebox{0.8\textwidth}{!}{
  \begin{tabular}{l l| c c c c c c }
    \toprule
    \textbf{Datasets}& \textbf{Metrics} & \textbf{$K=75$} & \textbf{$K=100$} & \textbf{$K=125$} & \textbf{$K=150$} & \textbf{$K=175$} & \textbf{$K=200$} \\
    \midrule
    \multirow{3}{*}{Pri-DDXplus}  
    & MC2 $\uparrow$ &  52.20 & 56.17 & 59.39 & 58.96 & \textbf{62.95} & 62.64 \\
    ~ & LLM-J $\uparrow$ & 4.51 & 4.38 & 4.45 & 4.33 & \textbf{4.71} & 4.55  \\
    ~ & TP $\uparrow$ & \textbf{24.31} & 21.51 & 19.72 & 20.07 & 22.68 & 21.91  \\
    \cmidrule(lr){2-8} 
     \multirow{3}{*}{Pri-NLICE}  
    & MC2 $\uparrow$ & 37.15 & 51.01 & \textbf{58.97} & 51.89 & 58.11 & 58.45 \\
    ~ & LLM-J $\uparrow$ & 3.27 & 3.66 & \textbf{3.80} & 3.44 & 3.40 & 3.62  \\
    ~ & TP $\uparrow$& \textbf{20.05} & 19.14 & 18.23 & 16.08 & 15.89 & 15.48  \\
    \cmidrule(lr){2-8} 
    \multirow{3}{*}{Pri-SLJA}  
    & MC2 $\uparrow$ & 28.75 & 30.65 & 35.07 & 32.41 & \textbf{35.13} & 32.08 \\
    ~ & LLM-J $\uparrow$& 5.21 & \textbf{5.41} & 5.00 & 5.33 & 5.15 & 5.28  \\
    ~ & TP $\uparrow$& \textbf{36.28} & 35.25 & 34.62 & 32.97 & 30.51 & 29.87 \\
    \toprule
  \end{tabular}
  }
  \caption{The performance of PrivacyRestore on the development set using various  numbers of edited heads $K$. 
  MC2 reflects classification capability, while LLM-J indicates generation performance.
  The TP assesses inference efficiency.
  We report results across three datasets to identify the optimal $K$ for each datasets. 
  The best results are highlighted in bold.}
   \label{tbl:vary_heads}
\end{table*}

\subsection{Varying LLM Backbone}
\label{app:v_model}
We evaluate the performance of PrivacyRestore and other privacy-preserving baselines on a larger model, Llama-13b-chat. 

As shown in Figure \ref{fig:13b}, 
PrivacyRestore outperforms the other baselines in terms of both MC2 and LLM-J values across all three privacy-preserving datasets. 
Notably, the performance of all privacy-preserving methods on the larger model, Llama-13b-chat, is worse than on the smaller model, Llama-7b-chat. 
This suggests that as model size increases, the model becomes more sensitive to the injected disturbances introduced by these privacy-preserving methods, leading to performance degradation.

    
    
  

\section{Inference Efficiency Analysis}
\label{app:inference_efficiency}
In this section, we provide a detailed analysis of the computational efficiency of our proposed method, specifically addressing its performance in terms of training time and inference throughput. Concerns regarding the computing resource requirements and time costs associated with training recovery vectors, building meta vectors, and other operations in practical applications have been noted. To address these, we have conducted a series of experiments to evaluate the efficiency of our approach, with a particular focus on its suitability for large-scale deployments.

Table~\ref{tab:inference_efficiency} presents a comprehensive overview of the performance metrics. The evaluation was conducted on three distinct datasets: Pri-DDXplus, Pri-NLICE, and Pri-SLJA. We report the total training time, the number of trainable parameters specific to our method, the ratio of these trainable parameters to the full model parameters, the inference throughput achieved using our method, the initial inference throughput (baseline), and the percentage of our method's inference throughput relative to the initial throughput.

\begin{table*}[!htbp]
  \centering
  \resizebox{0.8\textwidth}{!}{%
  \begin{tabular}{@{}lcccccc@{}}
    \toprule
    Dataset & \begin{tabular}[c]{@{}c@{}}Train \\ Time\end{tabular} & \begin{tabular}[c]{@{}c@{}}PrivacyRestore \\ Params\end{tabular} & \begin{tabular}[c]{@{}c@{}}Params \\ Ratio\end{tabular} & \begin{tabular}[c]{@{}c@{}}PrivacyRestore \\ TP\end{tabular} & \begin{tabular}[c]{@{}c@{}}Initial \\ TP\end{tabular} & \begin{tabular}[c]{@{}c@{}}TP \\ Ratio\end{tabular} \\
    \midrule
    Pri-DDXplus & 8h  & 26M & 4\% & 26.09 & 41.08 & 64\% \\
    Pri-NLICE   & 7h  & 8M  & 1\% & 32.33 & 41.44 & 78\% \\
    Pri-SLJA    & 10h & 26M & 4\% & 30.73 & 39.49 & 77\% \\
    \bottomrule
  \end{tabular}%
  }
  \caption{Experimental Results on Inference Efficiency of PrivacyRestore under different datasets. The table shows training time, trainable parameters (and their ratio to full model), inference throughput of our method, initial throughput, and the ratio of our method's throughput to the initial one.}
  \label{tab:inference_efficiency}
\end{table*}

The experimental results demonstrate that our method maintains a low computational overhead during the training phase. For instance, the training times for Pri-DDXplus, Pri-NLICE, and Pri-SLJA were 8 hours, 7 hours, and 10 hours, respectively. These durations are considered acceptable, particularly given that our approach focuses on training only the restoration vectors. Crucially, these restoration vectors constitute a small fraction of the total model parameters, ranging from just 1\% (for Pri-NLICE with 8M trainable parameters) to 4\% (for Pri-DDXplus and Pri-SLJA with 26M trainable parameters). This targeted training strategy significantly reduces the computational burden compared to retraining an entire model, making it highly efficient.

In the inference stage, our method demonstrates commendable performance by retaining a substantial portion of the original inference throughput. Specifically, the inference throughput achieved by our method was 26.09 for Pri-DDXplus, 32.33 for Pri-NLICE, and 30.73 for Pri-SLJA. When compared to the initial inference throughputs of 41.08, 41.44, and 39.49, respectively, our method sustains between 64\% and 78\% of the original throughput. This indicates that while introducing the restoration mechanism, the impact on inference speed is managed effectively. For example, with Pri-NLICE, our method achieved 78\% of the initial throughput while only requiring the training of 1\% of the model parameters. Similarly, for Pri-SLJA and Pri-DDXplus, we achieved 77\% and 64\% of the initial throughput, respectively. Furthermore, the method retains a high percentage (65\%-80\%) of the original inference throughput, indicating minimal overhead during the inference phase. These characteristics collectively demonstrate that our approach is not only effective but also computationally efficient, rendering it well-suited and feasible for deployment in large-scale applications where both training and inference costs are critical considerations. The results affirm that the method remains low-cost and efficient across both training and inference stages.

\section{Example Outputs of PrivacyRestore}
\label{app:example}
We provide some example outputs of our method in Figure \ref{fig:example}.
As shown in these examples, applying $d_\chi$-privacy to privacy spans results in outputs with higher ROUGE-L scores but lower LLM-J scores compared to our method.
After analyzing these outputs in detail, the high ROUGE-L scores from $d_\chi$-privacy on privacy spans likely result from a greater overlap with the initial output. 
However, the overlapping sections consist mainly of meaningless sentence structures and lack diagnostic information. 
Moreover, the final diagnosis is incorrect, leading to lower LLM-J scores. 
In contrast, PrivacyRestore generates outputs with a different structure but provides the same, correct diagnosis. 
As a result, our method achieves slightly lower ROUGE-L scores but significantly higher LLM-J scores compared to $d_\chi$-privacy on privacy spans.


\section{Privacy Spans Over-Removal}
\label{app:more_spans_remove}

To highlight the efficiency of PrivacyRestore in mitigating the impact of user errors during privacy span removal, we conducted a series of experiments to evaluate its robustness under adverse conditions. Recognizing that users may inadvertently remove longer spans than necessary, our study simulated scenarios where, in addition to the essential privacy spans, an extra 1\%, 5\%, 10\%, 20\%, 30\%, and even 50\% of non-privacy text was removed from the Pri-DDXPlus dataset.

Shown in Table \ref{table:over_removal}, although removing longer spans than necessary can impact performance, the degradation is minimal. Even with an extra 30\% of spans removed, our method still achieves robust scores—50.87 in MC1, 48.43 in MC2, 25.42 in ROUGE-L, and 3.65 in LLM-J.


\begin{table}[t]
\centering
\def\arraystretch{1.1} 
\resizebox{1\columnwidth}{!}{
\large
\begin{tabular}{l ccc}
\hline
 \textbf{Percentage of Spans Removal} & \textbf{MC1} & \textbf{MC2} &  \textbf{LLM-J}\\ 
\hline
 No More Spans Removed	&62.97	&60.19	&27.24 \\
1\% More Spans Removed	&62.13	&59.51	&27.21\\
5\% More Spans Removed	&59.13	&55.05	&27.16\\
10\% More Spans Removed	&57.45	&54.85&	26.51\\
\textbf{30\% More Spans Removed}	&50.87	&48.43	&25.42\\
50\% More Spans Removed	&42.47	&41.73&	23.90\\
\bottomrule
\end{tabular}
}
\caption{Performance of PrivacyRestore under varying percentages of additional span removal. The table displays the MC1, MC2, and LLM-J scores across varying percentages of removed non-privacy spans.}
\label{table:over_removal}
\end{table}

\begin{figure*}[!htbp]
  \centering
  \hspace{1em}
    \begin{minipage}{0.6\textwidth}
    
    
    \includegraphics[width=1\textwidth]{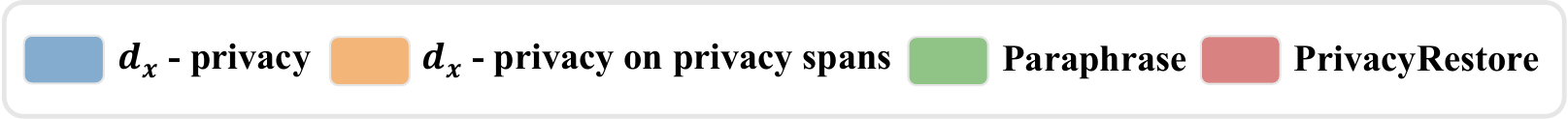}
    \label{fig:vmodel_legend}
  \end{minipage}
  \vspace{-2.5em}
  \vskip\baselineskip 
\hspace{-4em}
  \begin{minipage}{0.35\textwidth}
    \centering
    \subfigure[MC scores]{
        \includegraphics[width=1\textwidth]
        {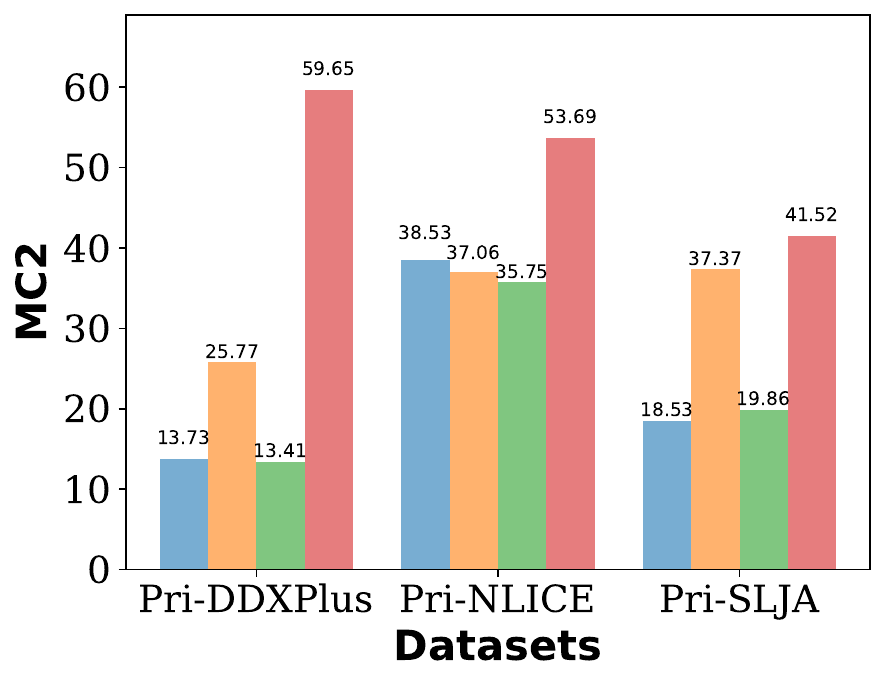}
        \label{fig:vmodel_mc}
    }
  \end{minipage}%
  \hspace{1em}
  \begin{minipage}{0.35\textwidth}
    \centering
    \subfigure[LLM-J scores]{
        \includegraphics[width=1\textwidth]
        {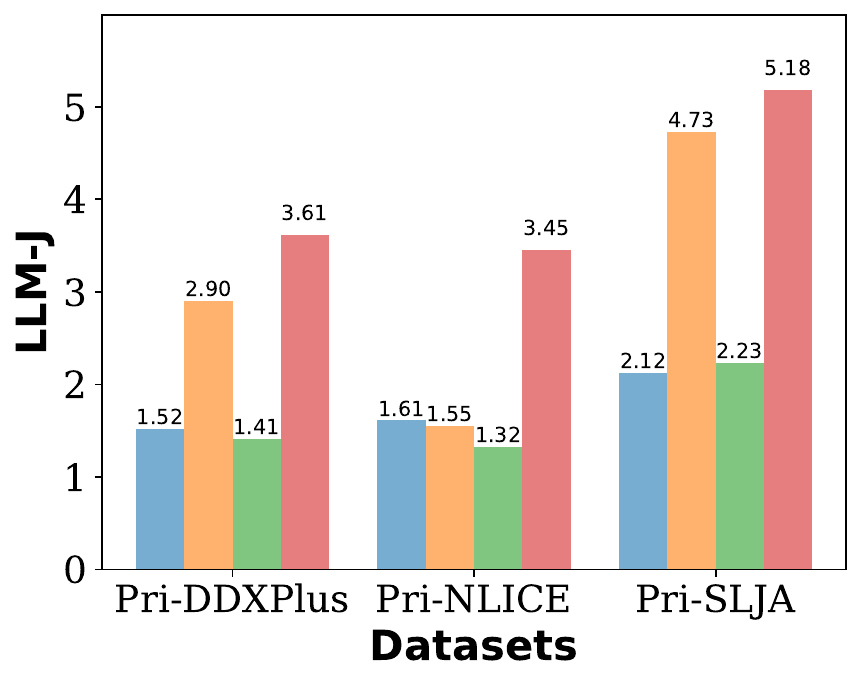}
        \label{fig:vmodel_gpt}
    }
  \end{minipage}
  
  \caption{The MC2 and LLM-J results of PrivacyRestore and other privacy-preserving baselines on larger model, Llama-13b-chat, across three datasets.}
  \label{fig:13b}
\end{figure*}
\newpage
\begin{figure*}[!htbp]
\centering 
    \vspace{-6pt}   
     \subfigure {  
    \includegraphics[width=0.85\textwidth]{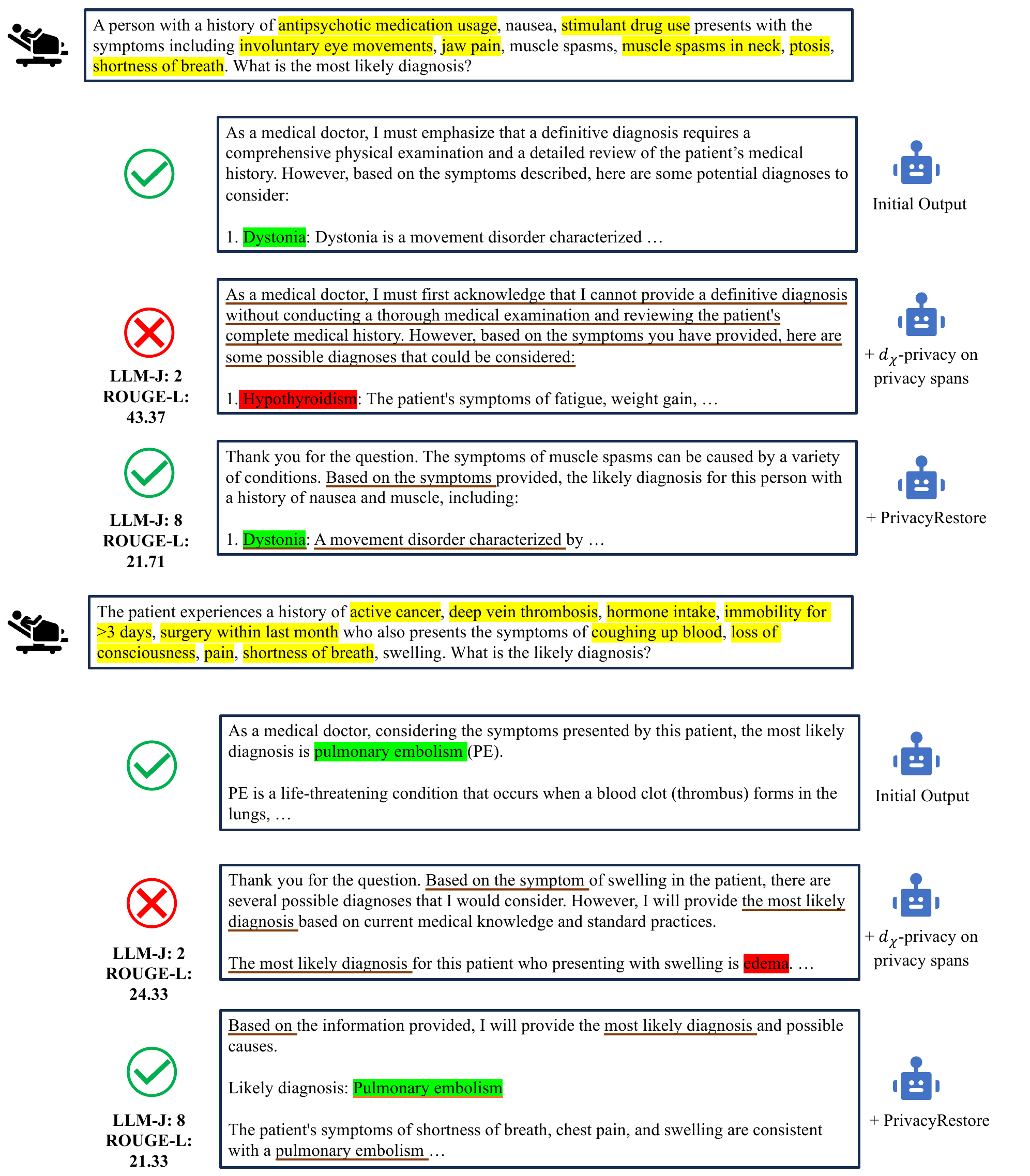}  
    }   
    \vspace{-6pt}
    \caption{
    Example Outputs of PrivacyRestore and $d_\chi$ on privacy spans in Pri-DDXPlus dataset.
    Text highlighted with a yellow background represents the privacy spans in user inputs.
    Text highlighted with a green background indicates the correct diagnosis.
    Text highlighted with a red background denotes the incorrect diagnosis.
    Underscored text marks sections that overlap with the initial output.
    }
    \vspace{-6pt}
    \label{fig:example}
\end{figure*}


\clearpage
\section{Prompt Template Details}
\subsection{Classification of Privacy Spans.}
\label{sec:class_privacy}
\subsubsection{Medical Datasets (Pri-DDXPlus/Pri-NLICE).}
Prompt template shown in Figure \ref{fig:med_assess_level} is for GPT and is used to classify symptoms in Pri-DDXPlus/Pri-NLICE dataset into sensitive and non-sensitive categories. 
GPT grades the symptoms on a scale of one to five based on sensitivity, with levels greater than three considered private spans in the Pri-DDXPlus/Pri-NLICE dataset.
\subsubsection{Legal Dataset (Pri-SLJA).}
Prompt template shown in Figure \ref{fig:legal_assess_level} is for GPT and is used to classify the case details in Pri-SLJA dataset into sensitive and non-sensitive categories. 
GPT grades the symptoms on a scale of one to five based on sensitivity, with levels greater than three considered private spans in the Pri-SLJA dataset.

\subsection{Rewriting of User Queries.}
\label{sec:rewrite_queries}
\subsubsection{Medical Datasets (Pri-DDXPlus/Pri-NLICE).}
The prompt template shown in Figure \ref{fig:rewrite_med} is designed for GPT and is utilized to rewrite medical queries in the Pri-DDXPlus and Pri-NLICE datasets.

\subsubsection{Legal Dataset (Pri-SLJA).}
The prompt template shown in Figure \ref{fig:rewrite_legal} is designed for GPT and is utilized to rewrite medical queries in the Pri-SLJA dataset.

\subsection{Generation Prompts.}
\label{sec:gene_prompt}
\subsubsection{Medical Datasets (Pri-DDXPlus/Pri-NLICE).}
Prompt template shown in Figure \ref{fig:gene_med} is for Llama model and is used during model generation for the Pri-DDXPlus/Pri-NLICE datasets.

\subsubsection{Legal Dataset (Pri-SLJA).}
Prompt template shown in Figure \ref{fig:gene_legal} is for Llama model and is used during model generation for the Pri-SLJA datasets.


\subsection{Evaluation of Generated Output.}
\label{sec:eval_gene}
\subsubsection{Medical Datasets (Pri-DDXPlus/Pri-NLICE).}
Prompt template shown in Figure \ref{fig:judge_med} is for GPT and evaluates the quality of generated output based on relevance, clarity, and accuracy, for Pri-DDXPlus/Pri-NLICE dataset. 
Scores range from 1 to 10, with higher values indicating better output.

\subsubsection{Legal Dataset (Pri-SLJA).}
Prompt template shown in Figure \ref{fig:judge_legal} is for GPT and evaluates the quality of generated output based on relevance, clarity, and accuracy, for Pri-SLJA dataset. 
Scores range from 1 to 10, with higher values indicating better output.

\subsection{Prompt Injection Attack.}
\label{app:prompt_PI}
Prompt template shown in Figure \ref{fig:prompt_PI} is for Llama model and is used to carry out a prompt injection attack, translating the garbled text back into the original text.

\begin{figure*}[!tp]
\centering 
     \includegraphics[width=0.99\textwidth]
     {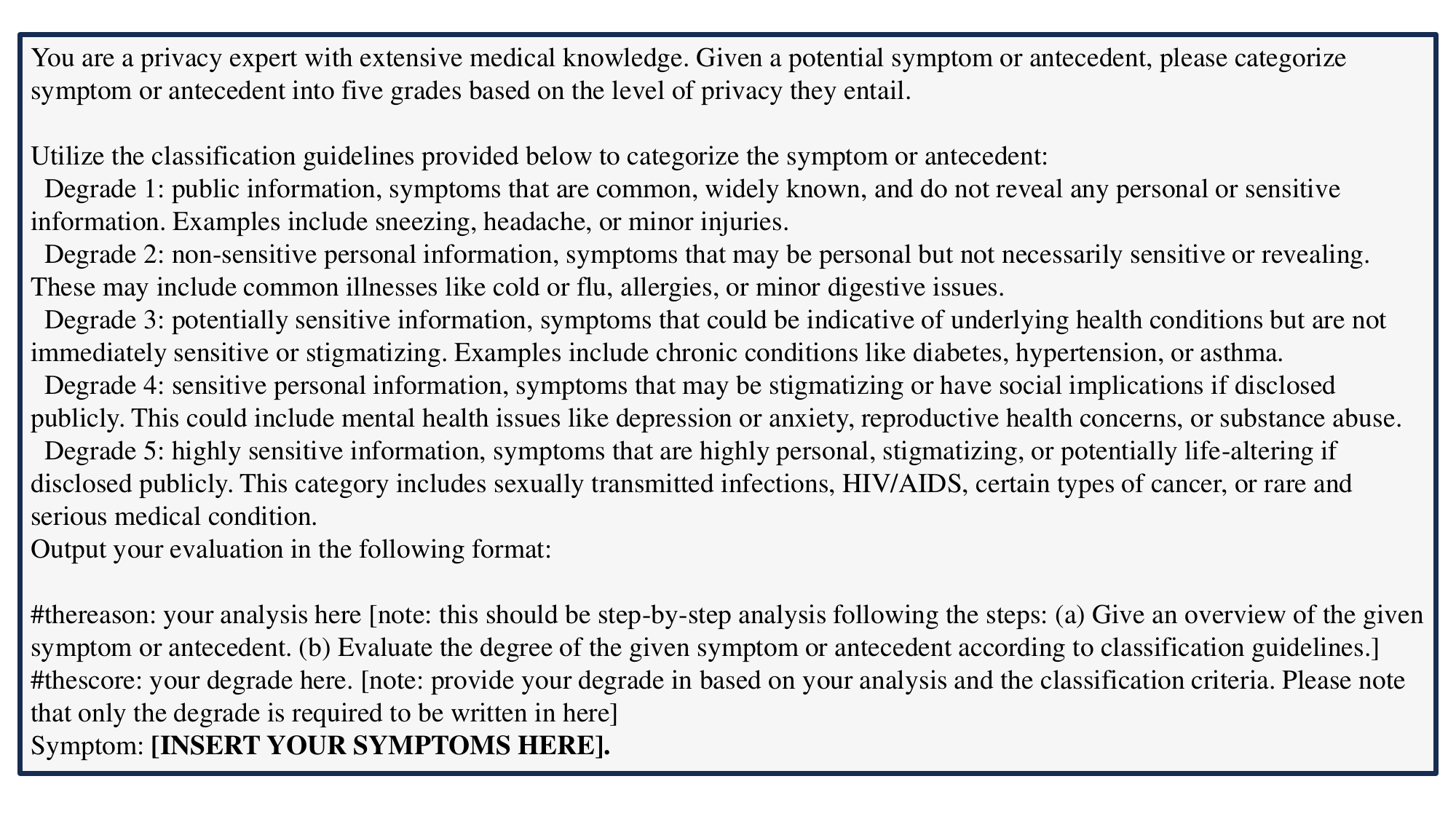}  
     \caption{Prompt template for classifying symptoms}
     \label{fig:med_assess_level}
\end{figure*}
\begin{figure*}[!t]
\centering 
     \includegraphics[width=0.99\textwidth]{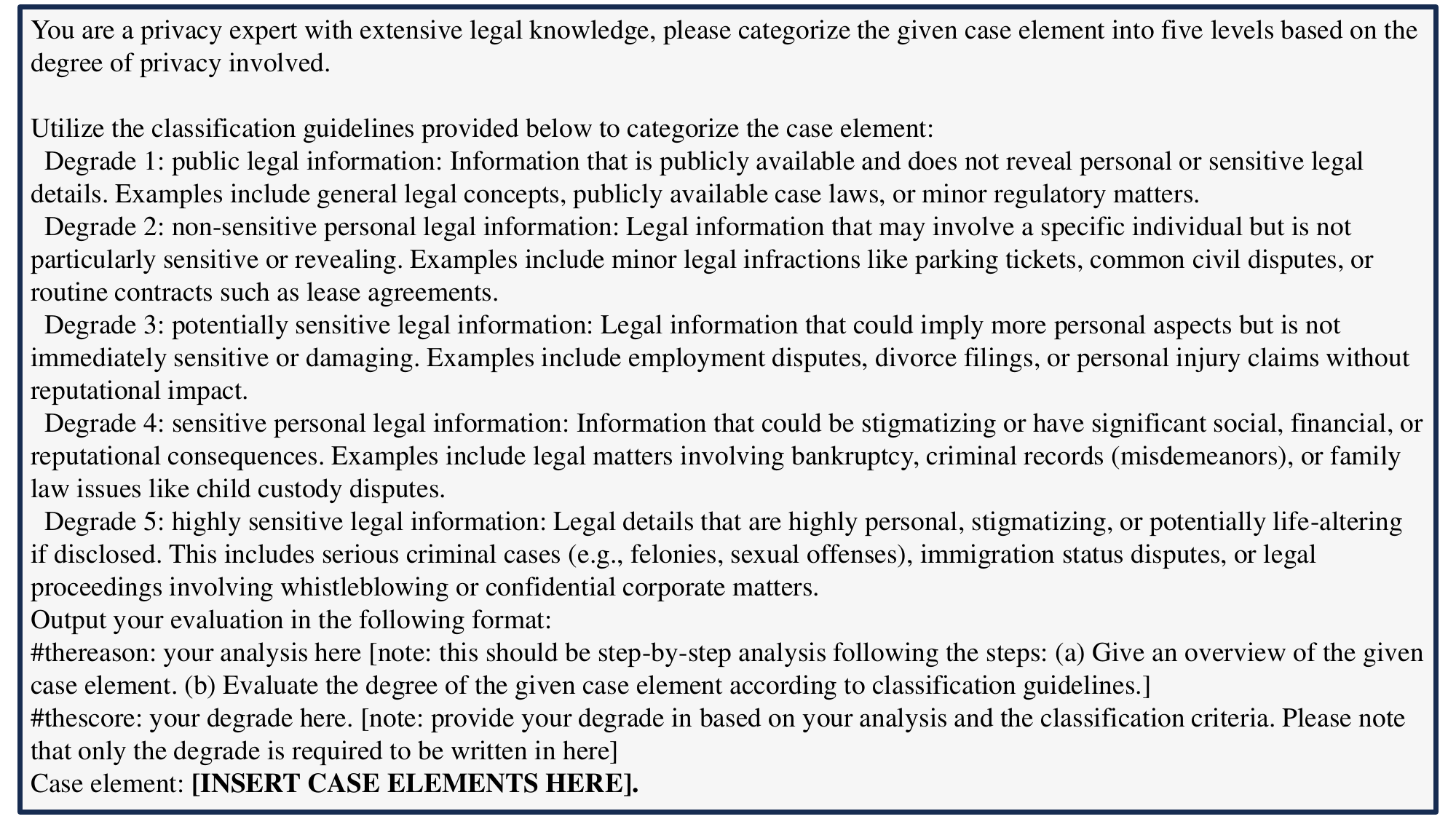}  
     \caption{Prompt template for classifying case details}
     \label{fig:legal_assess_level}
\end{figure*}

\begin{figure*}[!t]
\centering 
     \includegraphics[width=0.99\textwidth]{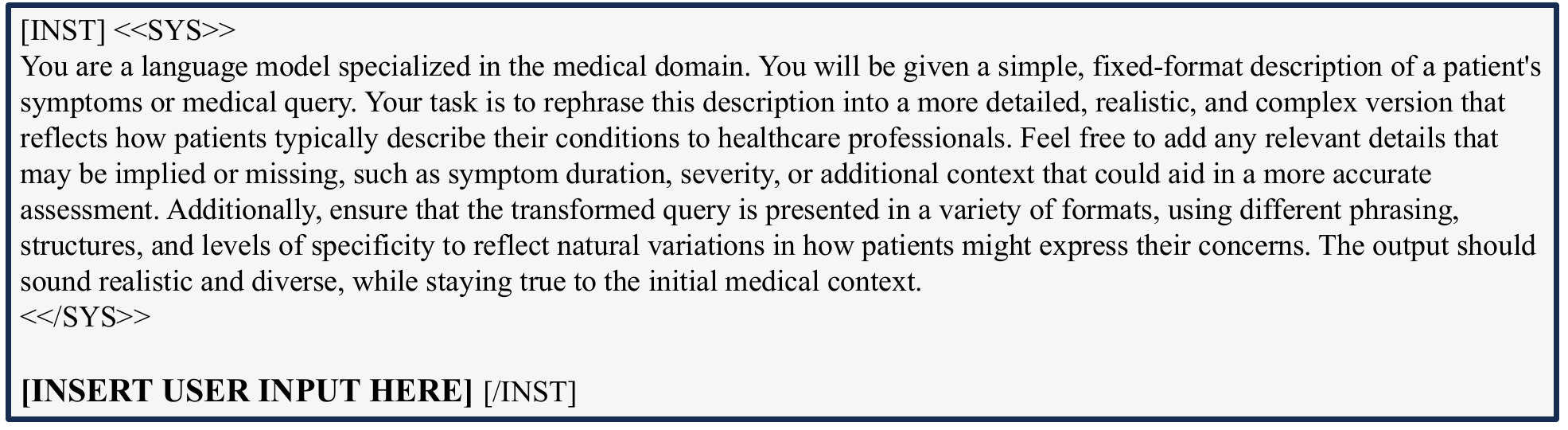}  
     \caption{Prompt template for rewriting queries in medical datasets}
     \label{fig:rewrite_med}
\end{figure*}
\begin{figure*}[!t]
\centering 
     \includegraphics[width=0.99\textwidth]{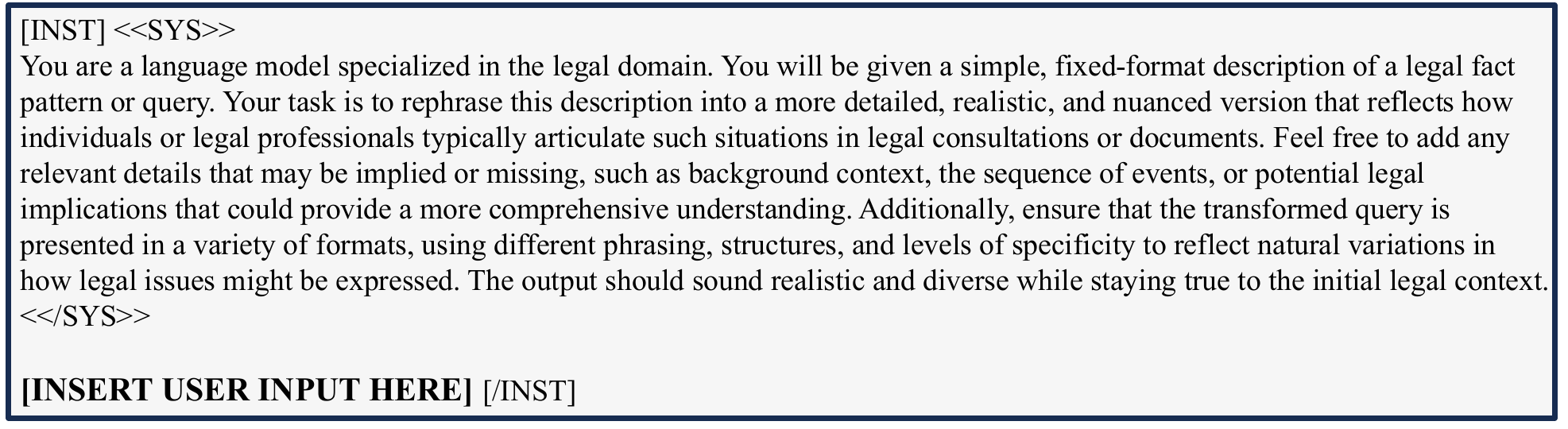}  
     \caption{Prompt template for rewriting queries in legal datasets}
     \label{fig:rewrite_legal}
\end{figure*}

\begin{figure*}[pt]
\centering 
     \includegraphics[width=0.99\textwidth]{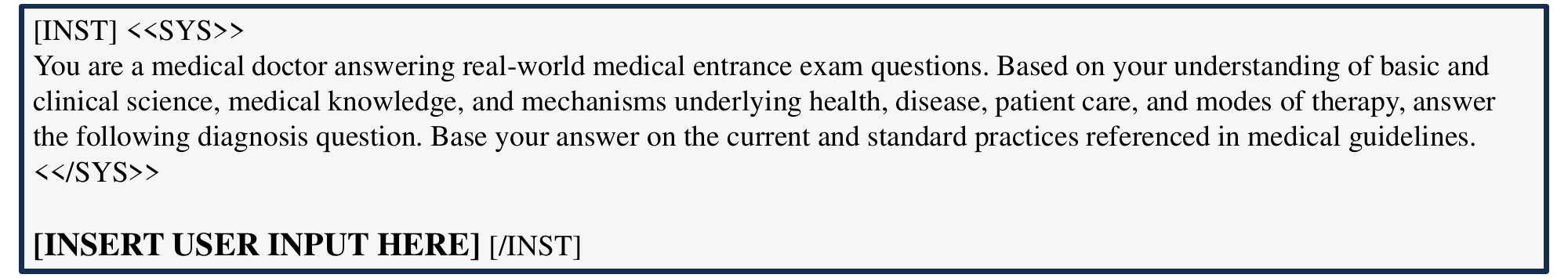}  
     \caption{Pri-DDXPlus/Pri-NLICE prompt template for model generation}
     \label{fig:gene_med}
\end{figure*}
\begin{figure*}[pt]
\centering 
     \includegraphics[width=0.99\textwidth]{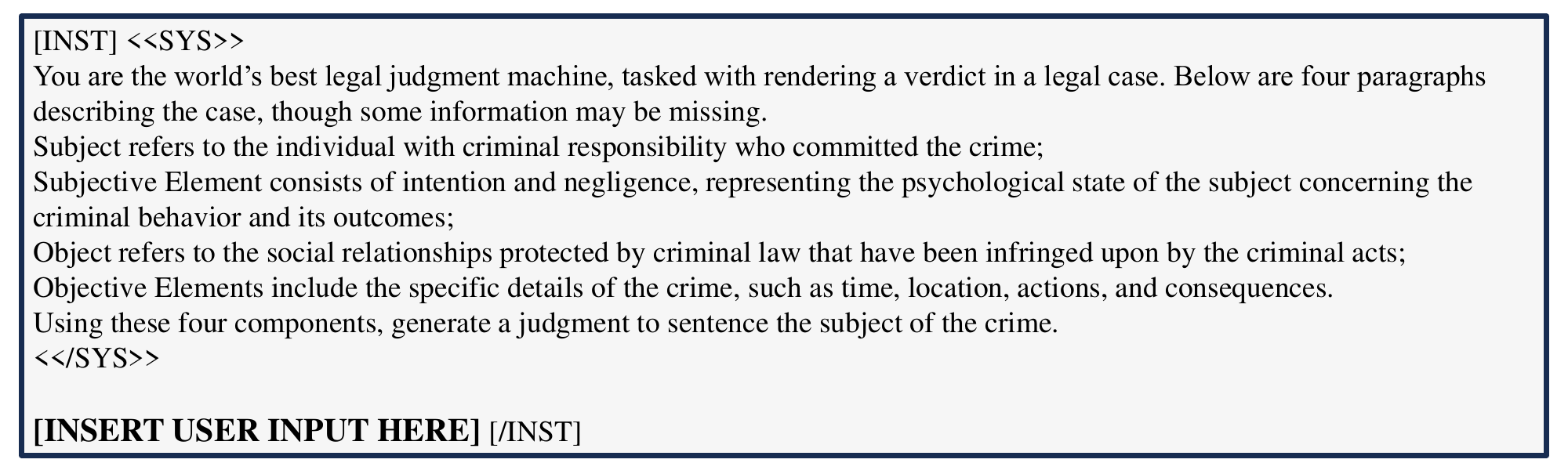}  
      \caption{Pri-SLJA prompt template for model generation}
     \label{fig:gene_legal}
\end{figure*}
\begin{figure*}[pt]
\centering 
     \includegraphics[width=0.99\textwidth]{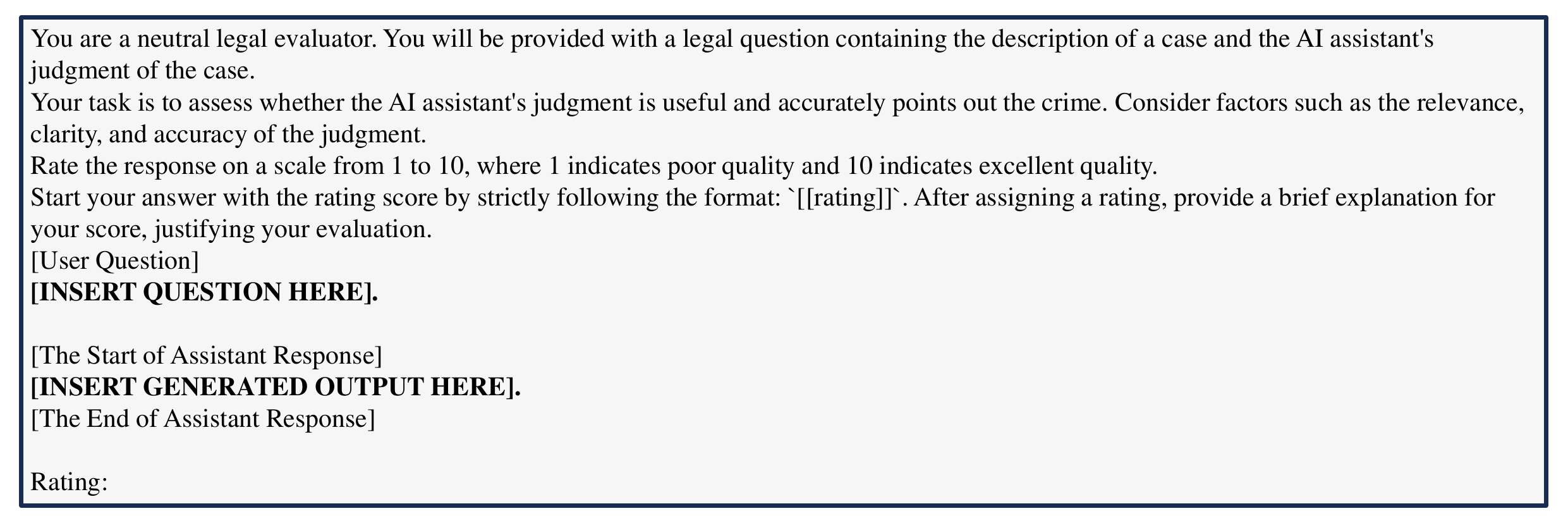}  
      \caption{Pri-DDXPlus/Pri-NLICE evaluation prompt template}
     \label{fig:judge_med}
\end{figure*}
\begin{figure*}[pt]
\centering 
     \includegraphics[width=0.99\textwidth]{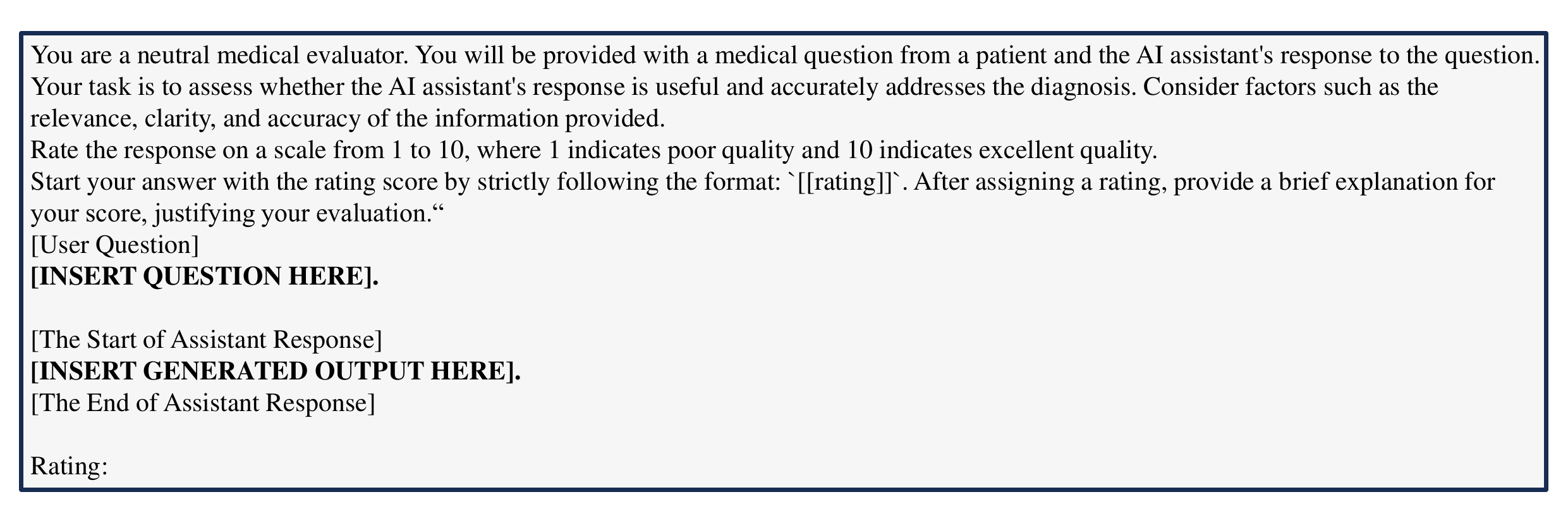}  
      \caption{Pri-SLJA evaluation prompt template}
     \label{fig:judge_legal}
\end{figure*}
\begin{figure*}[pt]
\centering 
     \includegraphics[width=0.99\textwidth]{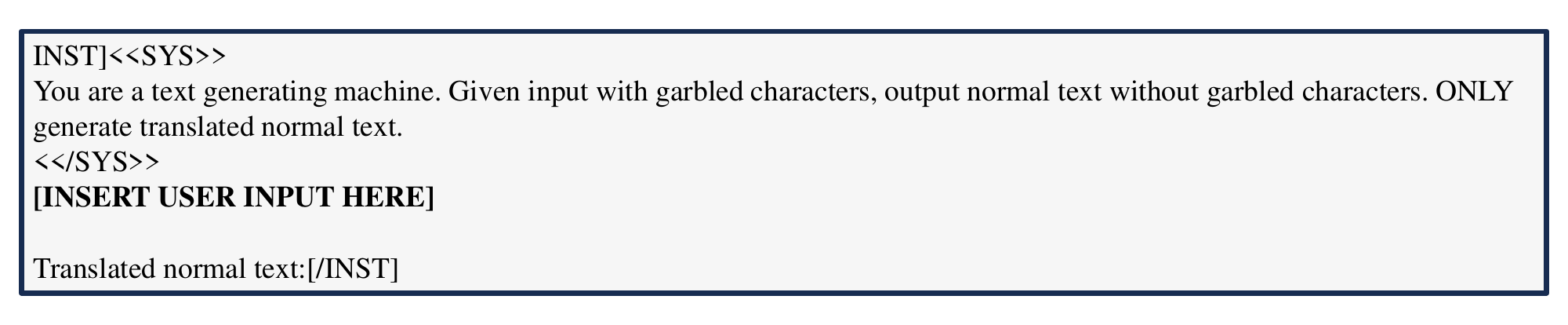}  
      \caption{Prompt injection attack template}
     \label{fig:prompt_PI}
\end{figure*}

\end{document}